\documentclass[pdflatex, referee, sn-vancouver, Numbered]{sn-jnl}

\usepackage{graphicx}%
\usepackage{multirow}%
\usepackage{amsmath,amssymb,amsfonts}%
\usepackage{amsthm}%
\usepackage{mathrsfs}%
\usepackage[title]{appendix}%
\usepackage{xcolor}%
\usepackage{textcomp}%
\usepackage{manyfoot}%
\usepackage{booktabs}%
\usepackage{algorithm}%
\usepackage{algorithmicx}%
\usepackage{algpseudocode}%
\usepackage{listings}%
\usepackage[switch]{lineno}
\usepackage{algorithmicx}
\usepackage{algpseudocode}

\begin{document}

\title[Article Title]{Modeling the 2022 Mpox Outbreak with a Mechanistic Network Model}


\author*[1]{\fnm{Emma G.} \sur{Crenshaw}}\email{emma\_crenshaw@g.harvard.edu}

\author[1]{\fnm{Jukka-Pekka} \sur{Onnela}}\email{onnela@hsph.harvard.edu}

\affil[1]{\orgdiv{Department of Biostatistics}, \orgname{Harvard TH Chan School of Public Health}, \orgaddress{\street{677 Huntington Avenue}, \city{Boston}, \postcode{02115}, \state{MA}, \country{USA}}}


 \abstract{\textbf{Background:} The 2022 outbreak of mpox affected more than 80,000 individuals worldwide, most of whom were men who have sex with men (MSM) who likely contracted the disease through close contact during sex. Given the unprecedented number of mpox infections and the new route of infection, there was substantial uncertainty about how best to manage the outbreak.
 
 \textbf{Methods:} We implemented a dynamic agent-based network model to simulate the spread of mpox in a United States-based MSM population. This model allowed us to implement data-informed dynamic network evolution to simulate realistic disease spreading and behavioral adaptations.
 
 \textbf{Results:} We found that behavior change, the reduction in one-time partnerships, and widespread vaccination are effective in preventing the transmission of mpox and that earlier intervention has a greater effect, even when only a high-risk portion of the population participates. With no intervention, 16\% of the population was infected (25th percentile, 75th percentiles of simulations: 15.3\%, 16.6\%). With vaccination and behavior change in only the 25\% of individuals most likely to have a one-time partner, cumulative infections were reduced by 30\%, or a total reduction in nearly 500 infections (mean: 11.3\%, $P_{25\%}$ and $P_{75\%}$: 9.6\%, 13.5\%). Earlier intervention further reduces cumulative infections; beginning vaccination a year before the outbreak results in only 5.5\% of men being infected, averting 950 infections or nearly 10\% of the total population in our model. We also show that sustained partnerships drive the early outbreak, while one-time partnerships drive transmission after the first initial weeks. The median effective reproductive number, $R_*^t$, at $t = 0$ days is 1.30 for casual partnerships, 1.00 for main, and 0.6 for one-time. By $t = 28$, the median $R_*^t$ for one-time partnerships has more than doubled to 1.48, while it decreased for casual and main partnerships: 0.46 and 0.29, respectively.
 
 \textbf{Conclusion:} With the ability to model individuals’ behavior, mechanistic networks are particularly well suited to studying sexually transmitted infections, the spread and control of which are often governed by individual-level action. Our results contribute valuable insights into the role of different interventions and relationship types in mpox transmission dynamics.}


\maketitle

\section{Background}\label{sec1}

Mpox is a vaccine-preventable viral infection that, before 2022, was primarily zoonotic and restricted to endemic areas of central and western Africa\cite{Bragazzi2023EpidemiologicalReview, Kwok2022Estimation2022}. However, in 2022, there was a significant global outbreak. Within a year, more than 87,000 cases were identified, approximately 30,000 of which were in the United States\cite{WHO20232022-23Trends}. These cases disproportionately affected gay, bisexual, and other men who have sex with men (MSM)\cite{WHO20232022-23Trends}. This outbreak marked a significant shift in transmission dynamics, with human-to-human transmission driving the epidemic. Transmission occurred primarily through contact with infectious lesions, with most cases being associated with sexual contact\cite{Patauner2022MonkeypoxPhysician, Charniga2024NowcastingLearned, Bragazzi2023EpidemiologicalReview}.

Due to the change in transmission modality from zoonotic to person-to-person, there was a great deal of clinical uncertainty about the transmissibility of the disease and uncertainty about how best to prevent future outbreaks. The concentration of cases among MSM  highlighted the importance of understanding mpox spread in sexual networks. There are two general frameworks of network modeling, statistical and mechanistic. Within each of these, the models can be either static or dynamic. Mechanistic network models, such as agent-based models, have been used previously in the study of HIV/AIDS and other sexually transmitted infections (STIs) in this population. These models are able to represent important aspects of sexual networks, such as repeated interactions with partners which cannot be captured without a network framework, and they can directly represent an individual's behavioral mechanisms, such as partnership formation, concurrency, or treatment and vaccination seeking\cite{Mei2010ComplexAmsterdam, Goodreau2017IsolatingStudy, Mattie2024ASpread, Jones2019ProportionAnalysis, Endo2022Heavy-tailed2022}. Thus, they are particularly well-suited to characterizing disease transmission governed by individual-level action, such as sexual contact patterns and partnership dynamics.

Recent modeling studies have focused on ascertaining epidemiological and clinical features of the disease, such as the transmission probability per sexual contact \cite{Chaturvedi2024EstimatingStudy, Kwok2022Estimation2022, Zhang2024TransmissionStudy}, the incubation period \cite{Charniga2022EstimatingOutbreak, Zhang2024TransmissionStudy, Miura2024TimeNetherlands}, and the basic reproductive number ($R_0$)\cite{Endo2022Heavy-tailed2022, Clay2024ModellingOutbreak, Zhang2024TransmissionStudy}. Focusing on the 2022 outbreak, there have also been studies looking at the relative importance of different interventions in ending the outbreak in the US and abroad\cite{Clay2024ModellingOutbreak, Chaturvedi2024EstimatingStudy, Delaney2022MorbidityStates, VanDijck2023TheStudy, Zhang2024TransmissionStudy, Banuet-Martinez2023Monkeypox:Insight}. However, a paucity of literature is available that synthesizes lessons learned from the 2022 outbreak and explores the range of outcomes with alternative interventions or intervention timings. Additionally, current network modeling approaches in the literature focus on either statistical network modeling using exponential random graphs\cite{Clay2024ModellingOutbreak, Charniga2024NowcastingLearned, VanDijck2023TheStudy, Spicknall2022Morbidity2022} or do not incorporate dynamic sexual partnerships\cite{Endo2022Heavy-tailed2022}.

 In this paper, we outline how we integrated data about the clinical course of mpox, sexual network characteristics of MSM, and individual- and policy-level interventions to generate a model of mpox transmission in a population of MSM in the United States. We also describe several simulations that compare the impact of the timing of interventions at different levels of compliance and intervention access to showcase the usefulness of these models as a way to inform individual behavior and health policy when the clinical features of an emerging outbreak are still uncertain.

\section{Methods}\label{sec2}

\subsection{Network Characterization}
We developed an agent-based model to simulate a dynamic sexual network of 10,000 MSM. The network is initialized as a configuration model with a degree distribution that is consistent with real-world sexual networks of MSM (Table~\ref{tab1})\cite{Le2024TemporalProcesses, Mei2010ComplexAmsterdam, Weiss2020EgocentricStudy, Newman2018Networks}. While the standard configuration model is limited to static networks, temporal variants of the model have also been recently proposed, such as in Le (2024)\cite{Le2024TemporalProcesses, Mei2010ComplexAmsterdam}. Our extension of the standard configuration model has two primary changes - we consider three different types of network ties (main, casual, and one-time partnerships), and we allow the graph to evolve over time to better represent the dynamic nature of sexual relationships. We chose network parameters to match empirical network structures reported in observational studies of MSM in the United States \cite{Weiss2020EgocentricStudy, Jenness2016ImpactFor}. Primarily, we use empirically defined population-level proportions of individuals in different relationship types, defined as the number of main and casual partners an individual will have concurrently, and population-level heterogeneity in the daily probability that an individual will have one-time contact. The network does not assume assortativity based on node features, such as preferred relationship type or sexual activity risk group; however, nodes seeking more partnerships are more likely to partner with other high-degree nodes because they are over-represented in the pool of potential partners seeking relationships.

\begin{figure}[H]
\centering
\includegraphics[width = 1\textwidth]{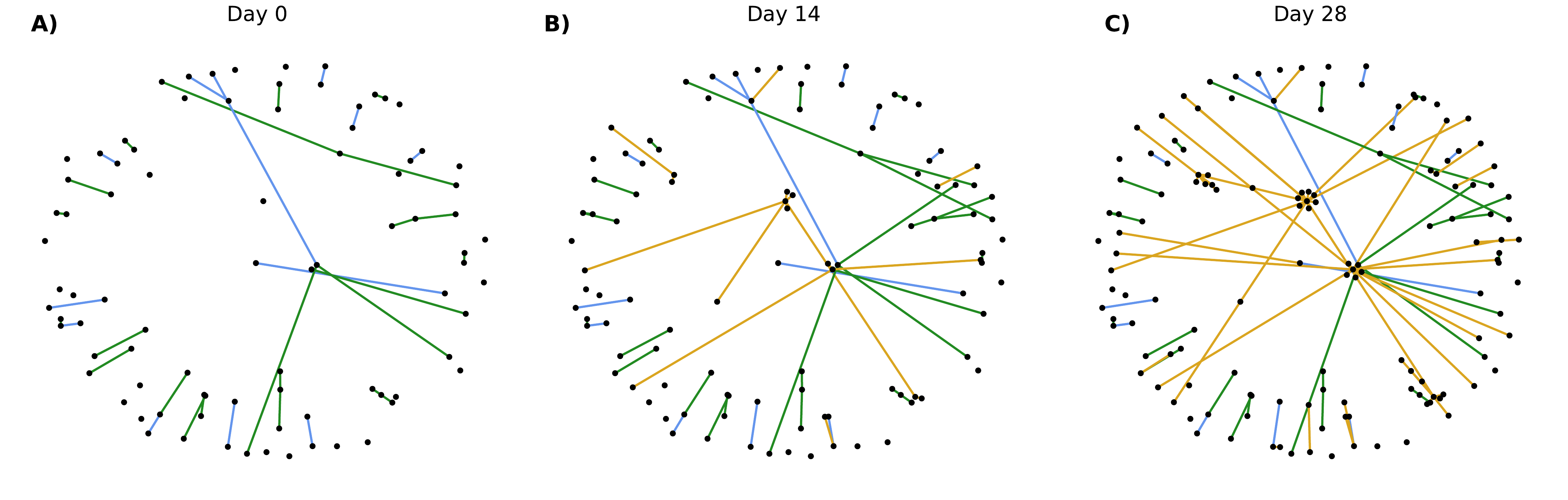}
\caption{\textbf{Cumulative edges for an ego-centric sample of 100 nodes and their alters}. Graphs show for day 0 (panel A), day 14 (Panel B), and day 28 (Panel C). Main partnerships are shown in blue, casual partnerships in green, and one-time partnerships in yellow. Nodes selected are treated as 'egos' and their partners as 'alters'. Thus, the only edges shown are those that involve an ego node. The number of main, casual, and one-time partnerships are 12, 23, 0 (Panel A); 12, 27, 15 (Panel B); 12, 27, 54 (Panel C)}\label{cumuledge}
\end{figure}

To demonstrate how the simulated sexual networks evolve over time, we show snapshots of the cumulative edges for an ego-centric sample of 100 nodes formed over 28 days (Figure \ref{cumuledge}). In these network visualizations, the 100 nodes sampled uniformly at random are treated as egos, and their partners (alters) are added to the visualization if they form an edge with an ego. In addition to ego-ego ties, the ego-alter ties are shown. From these visualizations, the importance of one-time partnerships in connecting different parts of the network over time becomes apparent.

\subsubsection{Main and casual partnerships}

The numbers of main and casual partnerships in the network follow the proportions in Table~\ref{tab1}, based on estimates from a national survey of MSM distributed between 2017 and 2019\cite{Weiss2020EgocentricStudy}, supplemented by data from two Atlanta-based surveys\cite{Jenness2016ImpactFor}. Each node is assigned probabilistically to a combination of main/casual partnership counts so that the final population percentages are matched to the real-world estimates and maintains that throughout the simulation. The probability of sexual contact on a given day with a particular partner is defined as 22\% for main partners and 14\% for casual partners, as found in the empirical surveys\cite{Jenness2016ImpactFor, Weiss2020EgocentricStudy}. The duration of each main or casual partnership is randomly defined via a draw from a geometric distribution parameterized by the mean partnership duration from observational studies: 407 days for main partnerships and 166 for casual partnerships\cite{Weiss2020EgocentricStudy, Goodreau2017IsolatingStudy}. Once a given partnership has reached its randomly pre-determined duration, the edge is dissolved, and nodes are randomly rewired to other nodes with the same type of partnership stub ("half edge") available. Nodes cannot rewire to their most recent connection; if no other nodes are available, they will wait to form another edge until another partnership of the desired type is dissolved to create new connections to available partners. In the appendix, we have characterized the typical wait time for a node when rewiring. For a network of size $N = 10,000$, fewer than 1\% of nodes looking to form a new main partnership are not able to do so immediately, and fewer than 0.002\% of those looking to form a casual partnership are not able to do so immediately (Table \ref{TableS1_rewiring}). A more detailed description of the network structure and summaries of network statistics are also available in the appendix.

\subsubsection{One-time partnerships}
Nodes were also randomly assigned to a stratum of sexual activity, defined as their daily probability of forming a one-time partnership, which is constant throughout the simulation. These probabilities are based on the same studies as the parameters for main and casual partnerships\cite{Weiss2020EgocentricStudy, Jenness2016ImpactFor}. Each node has a 19\% probability of being assigned to strata 1-5 and a 5\% probability of being assigned to strata 6; in this way, the strata can be thought of as quintiles of risk of one-time partnership with the addition of a `high-risk' group. This assumption is consistent with other modeling studies and is based on surveys of MSM\cite{Spicknall2022Morbidity2022, Clay2024ModellingOutbreak, Hernandez-Romieu2015HeterogeneityMSM, Goodreau2017IsolatingStudy}. These probabilities range from 0, when an individual will never have a one-time partner, to 0.286 in the high-risk group, which corresponds to approximately 8 one-time partners per month on average. By definition, the duration of a one-time partnership is one time step, and the probability of sexual contact is one.

\subsection{Epidemic Parameters}
To model the spread of mpox on this network, we used a discrete-time stochastic susceptible-exposed-infected-recovered (SEIR) transmission model. Each day, the occurrence of sexual contact in an existing partnership is treated as an independent random event. The probability of disease transmission in serodiscordant pairs is defined as the product of the probability of sexual contact and the overall transmission probability. Infected individuals are randomly assigned a given amount of time that they will be in the exposed and infected states, according to the distributions in Table~\ref{tab1}. Mean values for the distributions were defined as the average estimated time for each state in accordance with existing literature\cite{Charniga2022EstimatingOutbreak, Kwok2022Estimation2022, Bragazzi2023EpidemiologicalReview, Thornhill2022Monkeypox2022, Angelo2022EpidemiologicalStudy, Tarin-Vicente2022ClinicalStudy}. Once an individual's time in the current state has expired, they move to the next state. Initially, 0.1\% of nodes are infected at time $t=0$. These nodes were selected randomly from the 25\% of nodes most likely to have a one-time sexual partnership, i.e., those in the top two strata of sexual activity, to ensure that it is unlikely the spreading process dies out by chance within the first few time steps.

During the 2022 outbreak, individuals were recommended to abstain from sexual contact until resolution of their symptoms. However, the severity of symptoms ranged, and it is possible that people may not have recognized that they were infected, particularly early in the syndromic stage, delaying isolation. This is especially true early in the outbreak, as people were less likely to be aware of the symptoms of mpox. To account for this uncertainty in our simulations, we assumed that 80\% of individuals would seek clinical care, allowing for heterogeneity in behavior, symptom severity, and access to healthcare. Once an individual was infected, we assumed a 15-day diagnosis delay at the outbreak's start\cite{VanDijck2023TheStudy}. This delay decreased by approximately one day every four days of the simulation to simulate increased awareness by individuals and healthcare providers as the outbreak progressed\cite{Zucker2023CROIMpox}. The minimum allowed delay was 5 days. Once diagnosed, the node is assumed to completely isolate from all partners. A sensitivity analysis in the supplement shows a less optimistic isolation scheme; instead of full isolation, nodes that isolate will avoid one-time partnerships but only decrease their probability of contact with main and casual partners by 50\%.

\subsection{Interventions}
We investigated several potential interventions at different levels of intensity to determine their effect on cumulative mpox incidence after 250 days. Based on the literature and reports of real-world interventions, we simulated individual behavior change and vaccination. Our main results are presented for vaccination, which begins at 30 days, and behavioral change, which begins at 70 days. These timings were selected to represent the real-world outbreak: the first mpox case was diagnosed on May 17, 2022, vaccines became available on May 22, 2022, and a survey of MSM reported that individuals were engaging in behavioral change was conducted on August 5-15, 2022\cite{CDC2022TechnicalCDC, CDC2022ImpactBehaviors}. Assuming that the virus was circulating two weeks shortly before the first case was diagnosed, the 30-day and 70-day timings reflect these real-world interventions. Parameters used in simulation are in Table~\ref{tab1}.

\subsubsection{Behavior change}
According to preliminary survey results, approximately 50\% of MSM in the US reduced their overall number of sexual partners, and 50\% reduced their number of one-time partners\cite{Delaney2022MorbidityStates}. We wanted to examine the results if all members of the network were to reduce their sexual activity compared to only those most likely to form a one-time partnership doing so. To investigate this, we simulated a 50\% reduction in all individuals' probability to form a one-time partnership compared to the same reduction in only sexual activity strata 5 and 6, the 25\% of individuals with the greatest probability of forming a one-time partnership. For example, for individuals in stratum 6, this would correspond to a decrease from 8 one-time partners per month on average to 4 one-time partners.

\subsubsection{Vaccination}
The vaccine approved in the US for the prevention of mpox, JYNNEOS, requires two doses 28 days apart\cite{Dalton2023Estimated2023}. Vaccine coverage is still relatively low; approximately 22.7\% of MSM have received a first dose as of January 2023 \cite{Owens2023JYNNEOS2023}. In our simulation, we modeled the availability of vaccines using the Center for Disease Control and Prevention's (CDC) report of the number of vaccinations administered per week in the US between May 22, 2022 and July 11, 2023. The average number of vaccines administered per day was adjusted for the population size of our network and used as the total number of vaccinations to administer each day of the simulation. Further details and the data used are available in the supplement. 

Given that the vaccine supply was limited at the beginning of the outbreak, we simulated two vaccination schemes: random vaccination and vaccination of those in the highest stratum of sexual activity. Additionally, we looked at the effect of fast-tracking or delaying vaccine introduction to determine the sensitivity of disease transmission to potential improvements in epidemic preparedness. 

\subsection{Analyses}
\begin{figure}[H]
\centering
\includegraphics[width = 1\textwidth]{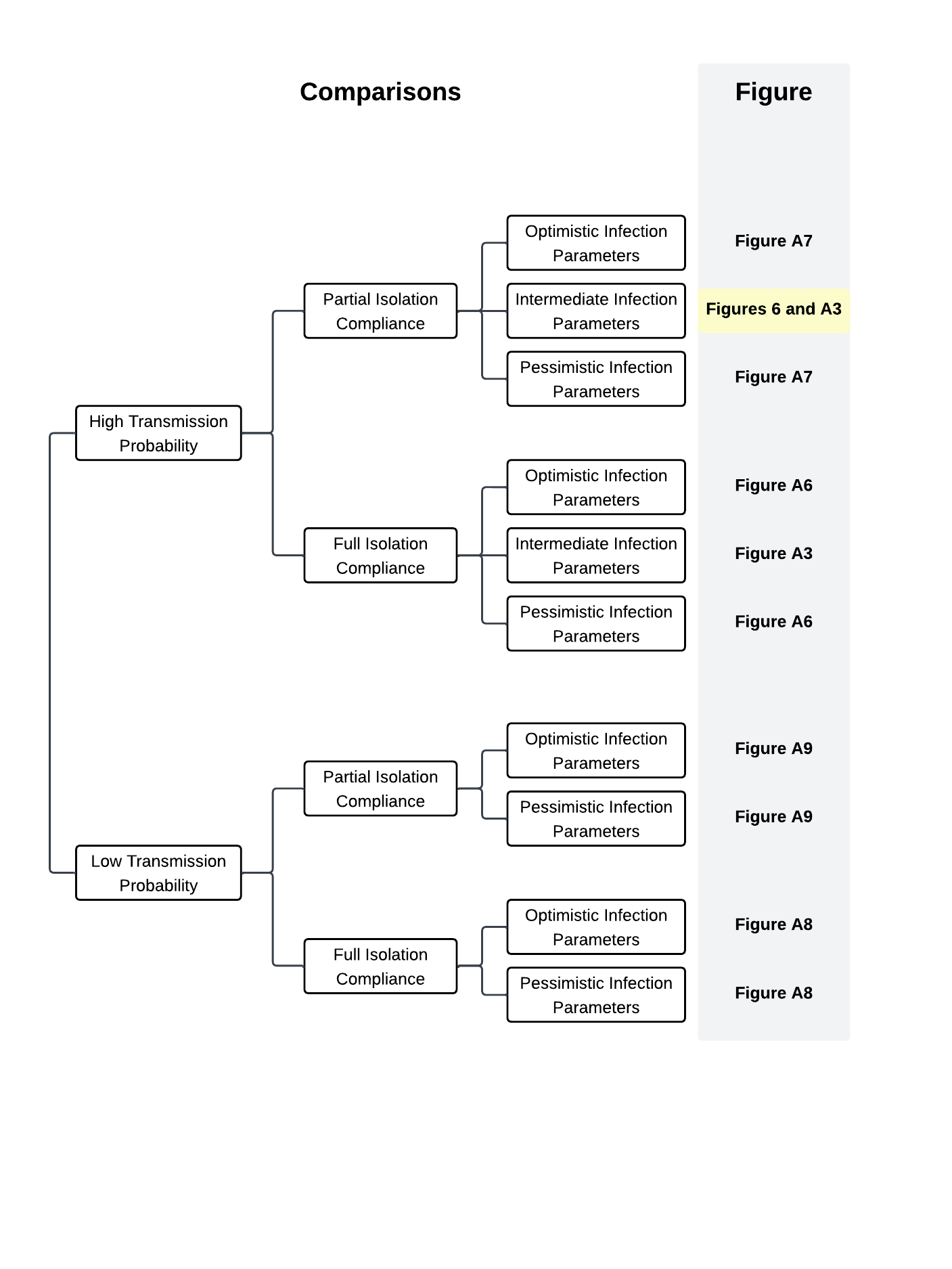}
\caption{\textbf{Explanation of Figures} Comparisons of different intervention timings and intensities under different simulation conditions can be found in the indicated figures. The primary comparison results are in Figure \ref{heatmap} and copied in Figure \ref{heatmap_both} to enhance comparability.}\label{navigate}
\end{figure}

All simulations and analyses were done in Python, and the code has been made publicly available\cite{CrenshawMpox-model}. Due to the stochastic nature of the model, wide variations in the simulations are possible. For example, it is possible that all infected individuals recover before infecting a partner, leading to the outbreak ending early. Similarly, given that the network is sparse, it is possible that the infection may become isolated within a particular portion of the network. Therefore, each analysis is the result of 100 independent simulations. Due to the complexity of a mechanistic network model, confidence intervals cannot be calculated; instead, we provide point-wise averages with the 25th and 75th percentiles of final cumulative infections to show the variability in simulation results. The effective reproductive number at time $t$ for a particular simulation, $R^t_*$, is calculated as the average number of infections secondary to any node that was infectious in the week before time $t$. At time $t=0$, this is equivalent to $R_0$, the basic reproduction number. Results are presented as the average $R^t_*$ over the 100 simulations. Due to the number of sensitivity analyses, we included Figure \ref{navigate} to show help readers navigate to figures showing the comparison of different intervention timings under different simulation conditions.

\begin{table}[h!]
\caption{Parameters for simulation scenarios}\label{tab1}%
\begin{tabular}{@{}lll@{}}
\toprule
\multicolumn{1}{l}{\textbf{Network Model}} \\
\midrule
$rt_k$ && Relationship type, percent of nodes \cite{Jenness2016ImpactFor, Weiss2020EgocentricStudy}\\
& 47.1&\quad \quad 0 main, 0 casual \\
& 16.7 &\quad \quad 0 main, 1 casual \\
& 7.4 &\quad \quad 0 main, 2 casual \\
& 22.0 &\quad \quad 1 main, 0 casual \\
& 4.7 &\quad \quad 1 main, 1 casual \\
& 2.1 &\quad \quad 1 main, 2 casual \\
$\pi_{o,k}$ &  & Daily probability of one-time partnership formation \cite{Jenness2016ImpactFor, Weiss2020EgocentricStudy}\\
 & 0 & \quad \quad stratum 1 \\
 & 0.001 & \quad \quad stratum 2 \\
 & 0.0054 & \quad \quad stratum 3 \\
 & 0.0101 & \quad \quad stratum 4 \\
 & 0.0315 & \quad \quad stratum 5 \\
 & 0.286 & \quad \quad stratum 6 \\
$n_{o}$ & Geometric($1 - \pi_{o,k}$) & Number of one-time contacts, per day \cite{Weiss2020EgocentricStudy}\\
$rd_{m,e}$ & Geometric(1/407) & Duration of main partnership, days \cite{Jenness2016ImpactFor, Weiss2020EgocentricStudy}\\
$rd_{c,e}$ & Geometric(1/166) & Duration of casual partnership, days \cite{Jenness2016ImpactFor, Weiss2020EgocentricStudy}\\
\toprule
\multicolumn{1}{l}{\textbf{Epidemic Model}} \\
\midrule
$\beta$ & 0.9 & Probability of transmission per sexual contact\\
$\pi_m$ & 0.22 & Daily probability of sexual contact, main partnership \cite{Jenness2016ImpactFor, Weiss2020EgocentricStudy}\\
$\pi_c$ & 0.14 & Daily probability of sexual contact, casual partnership \cite{Jenness2016ImpactFor, Weiss2020EgocentricStudy}\\
$t_{e,k}$ & Normal(7, 1) & Time spent in exposed state, days \cite{Charniga2022EstimatingOutbreak, Kwok2022Estimation2022, Bragazzi2023EpidemiologicalReview, Thornhill2022Monkeypox2022, Angelo2022EpidemiologicalStudy, Tarin-Vicente2022ClinicalStudy}\\
$t_{i,k}$ & Normal(27, 3) & Time spent in infectious state, days \cite{Charniga2022EstimatingOutbreak, Kwok2022Estimation2022, Bragazzi2023EpidemiologicalReview, Thornhill2022Monkeypox2022, Angelo2022EpidemiologicalStudy, Tarin-Vicente2022ClinicalStudy}\\
\toprule
\multicolumn{1}{l}{\textbf{Interventions}} \\
\midrule
$\pi_{b,k}$ & 0.5 & Decrease in probability of one-time partnership formation\\
$\pi_{v_1}$ & 0.14 & Probability of vaccination, only one dose \cite{Owens2023JYNNEOS2023}\\
$\pi_{v_2}$ & 0.227 & Probability of vaccination, two doses \cite{Owens2023JYNNEOS2023}\\
$VE_1$ & 0.358 & Vaccine efficacy, one dose \cite{Deputy2023VaccineStates}\\
$VE_2$ & 0.66 & Vaccine efficacy, both doses \cite{Deputy2023VaccineStates}\\
\botrule
\end{tabular}
\footnotetext{Node-specific attributes have the subscript '$X_{x,k}$', edge-specific attributes have the subscript '$X_{x,e}$}
\end{table}

\section{Results}
We present the results of several analyses, comparing the cumulative number of infections across different intervention types and timings. Our primary results show behavioral change beginning at day 70 of the simulation and vaccination beginning at day 30 and assume that individuals who are diagnosed comply with full isolation. Sensitivity analyses examining the impact of partial isolation compliance and more or less optimistic estimates of the disease-spreading process can be found in the appendix.

\subsection{Comparison of interventions}
\begin{figure}[H]
\centering
\includegraphics[width = 1\textwidth]{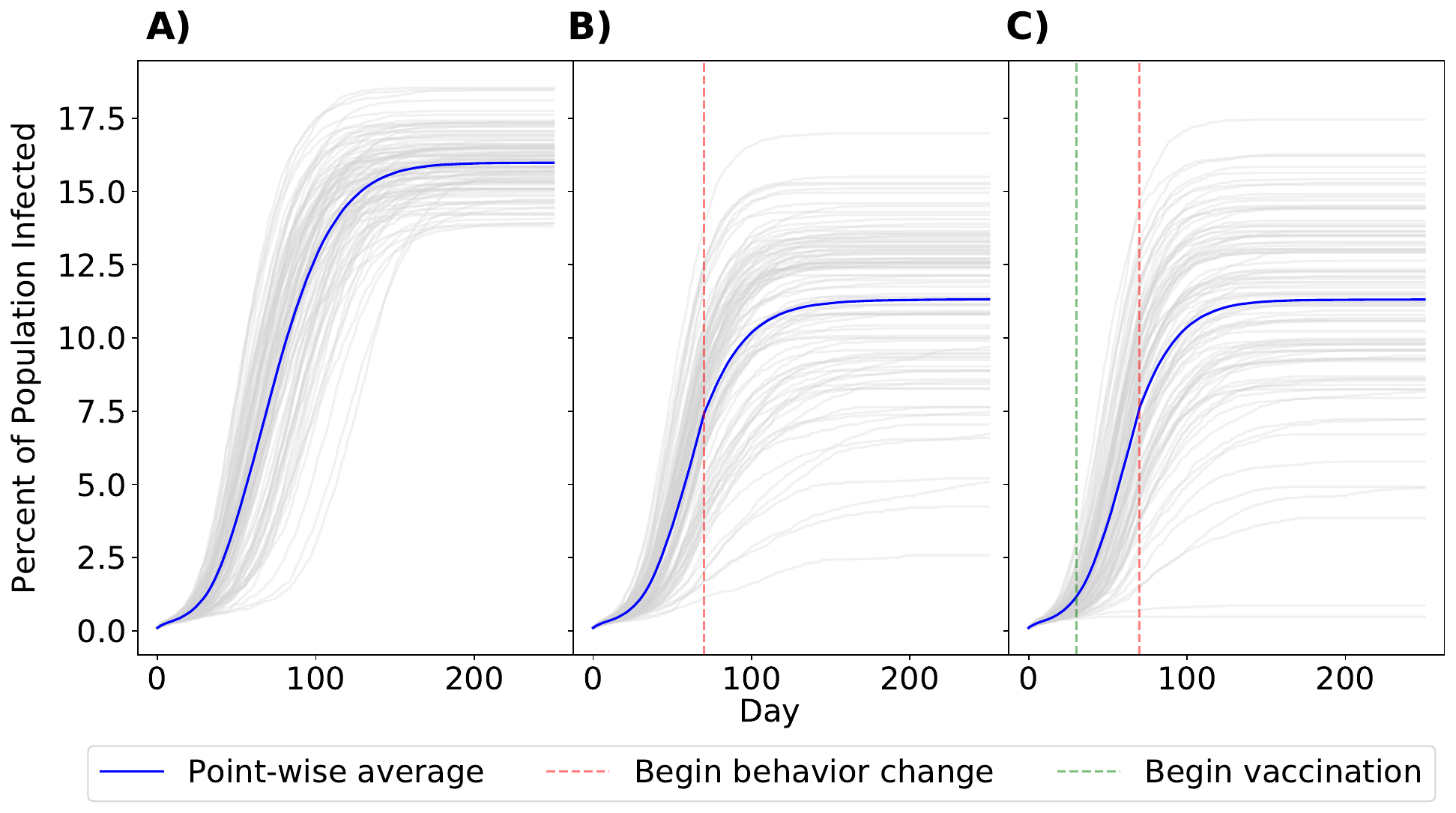}
\caption{\textbf{Comparison of universal interventions.} Panels indicate the percent of network infected with mpox after 250 days with no intervention (Panel A), universal behavior change (Panel B), or universal behavior change with vaccination (Panel C). Grey lines denote individual simulations. The point-wise average is shown in blue. Vertical lines indicate the day of intervention initiation.}\label{compint}
\end{figure}

Our baseline model simulates an epidemic where no personal intervention or policy is in place to prevent the spread of mpox. In this scenario, we see a rapid increase in cases until more than 15\% of the population of MSM is infected (mean percent of population infected: 15.98\%, 25th and 75th percentiles of infections:  15.31\%, 16.58\%). To get a better understanding of the impact of behavior change and vaccination in preventing infection, we modeled universal behavior change, a 50\% reduction in the probability of having a one-time partner, beginning at day 70 of our simulation and universal availability of vaccination beginning at day 30 (Figure \ref{compint}). From behavior change alone, we see a 30\% reduction in the total number of cases (mean: 11.32\%, $P_{25\%}$ and $P_{75\%}$: 9.56\%, 13.11\%), translating to approximately 500 averted infections. Adding a universal availability of vaccination does not meaningfully affect the results (mean: 11.31\% $P_{25\%}$ and $P_{75\%}$: 9.59\%, 13.51\%).

\begin{figure}[H]
\centering
\includegraphics[width = 1\textwidth]{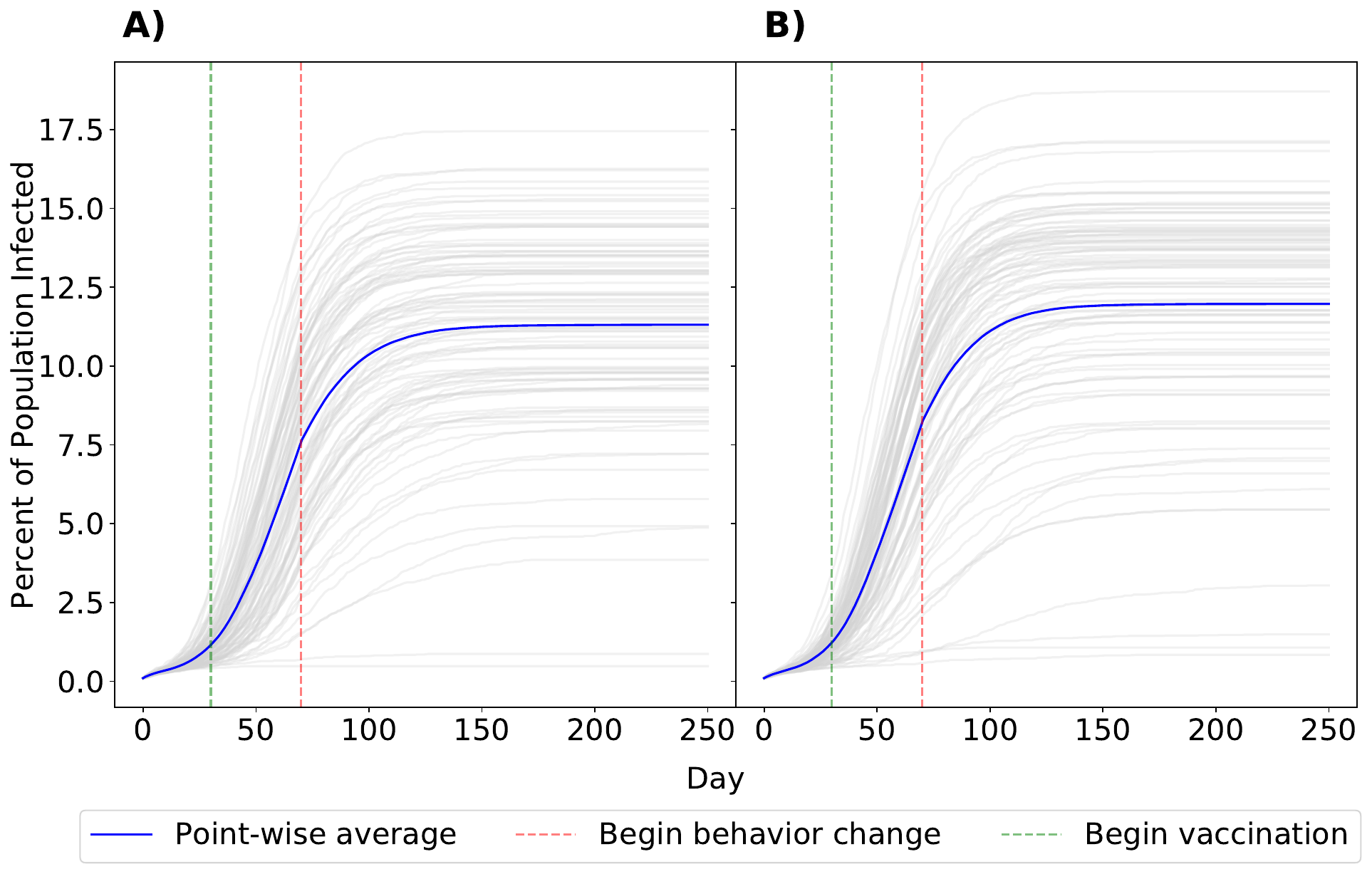}
\caption{\textbf{Comparison of scenarios with universal or targeted intervention.} Panels indicate the percent of network infected with mpox after 250 days with universal behavior change with vaccination (Panel A) or intervention only in the 25\% of men most likely to have a one-time partner (Panel B). Grey lines denote individual simulations. The point-wise average is shown in blue. Vertical lines indicate the day of intervention initiation.}\label{univtarget}
\end{figure}

\begin{figure}[H]
\centering
\includegraphics[width = 1\textwidth]{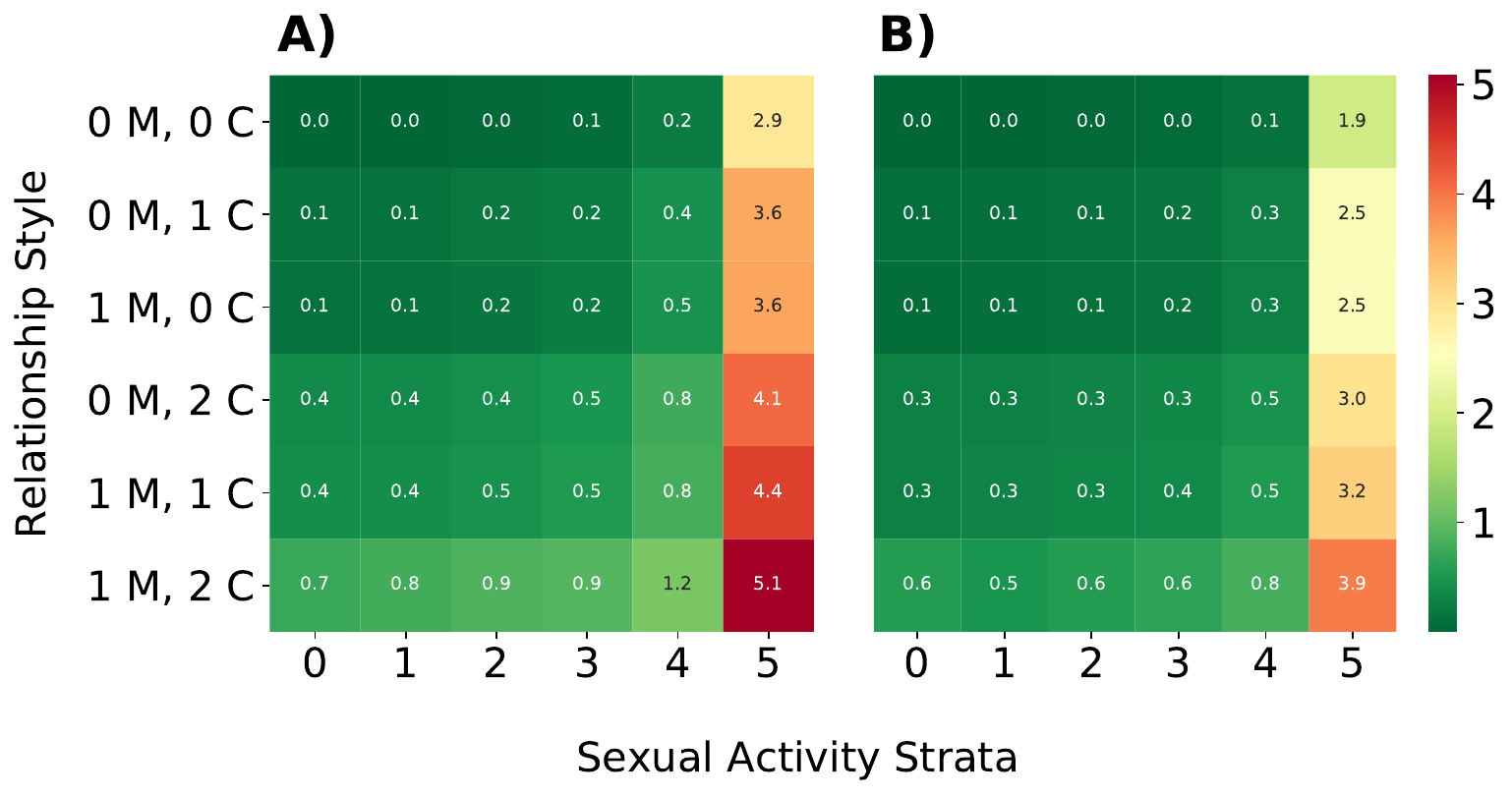}
\caption{\textbf{Comparison of average number of at-risk sexual interactions by relationship type and sexual activity strata} Cell values indicate the average number of serodiscordant sexual interactions individuals of a given relationship type and sexual activity stratum have over 250 days with no intervention (Panel A) or intervention only in the 25\% of men most likely to have a one-time partner (Panel B). Rows indicate relationship type (preferred number of main (M) and casual (C) partners), while columns indicate sexual activity strata, or a node's daily probability of having a one-time partner.}\label{contacts_10k}
\end{figure}

Given that universal behavior change is unlikely, we also examined a scenario with a targeted intervention, such that only men with the highest-risk sexual behavior (those in strata 5 and 6 of the daily probability of having a one-time partner) changed their behavior and received vaccination (Figure \ref{univtarget}). In this simulation, we saw minimal impact on intervention efficacy (mean: 11.97\% $P_{25\%}$ and $P_{75\%}$: 10.39\%, 14.23\%), indicating that the majority of prevented cases are due to behavior change by individuals most likely to form one-time partnerships. The effect of this targeted intervention can also be seen in infection risk, or the number of sexual interactions a node has as part of a serodiscordant partnership (Figure \ref{contacts_10k}). While the number of at-risk sexual contacts does increase with a node's degree, or number of partners, it is most affected by whether a node is in the highest-risk sexual behavior group (those in strata 6 of sexual activity, and to a lesser degree, strata 5). However, implementing behavior change decreases the average number of at-risk sexual contacts these nodes have over the course of the epidemic, lowering their risk of infection.

Sensitivity analyses examining reduced compliance to isolation from main and casual partnerships are available in the appendix.

\subsection{Comparison of intervention timings}

\begin{figure}[H]
\centering
\includegraphics[width = 1\textwidth]{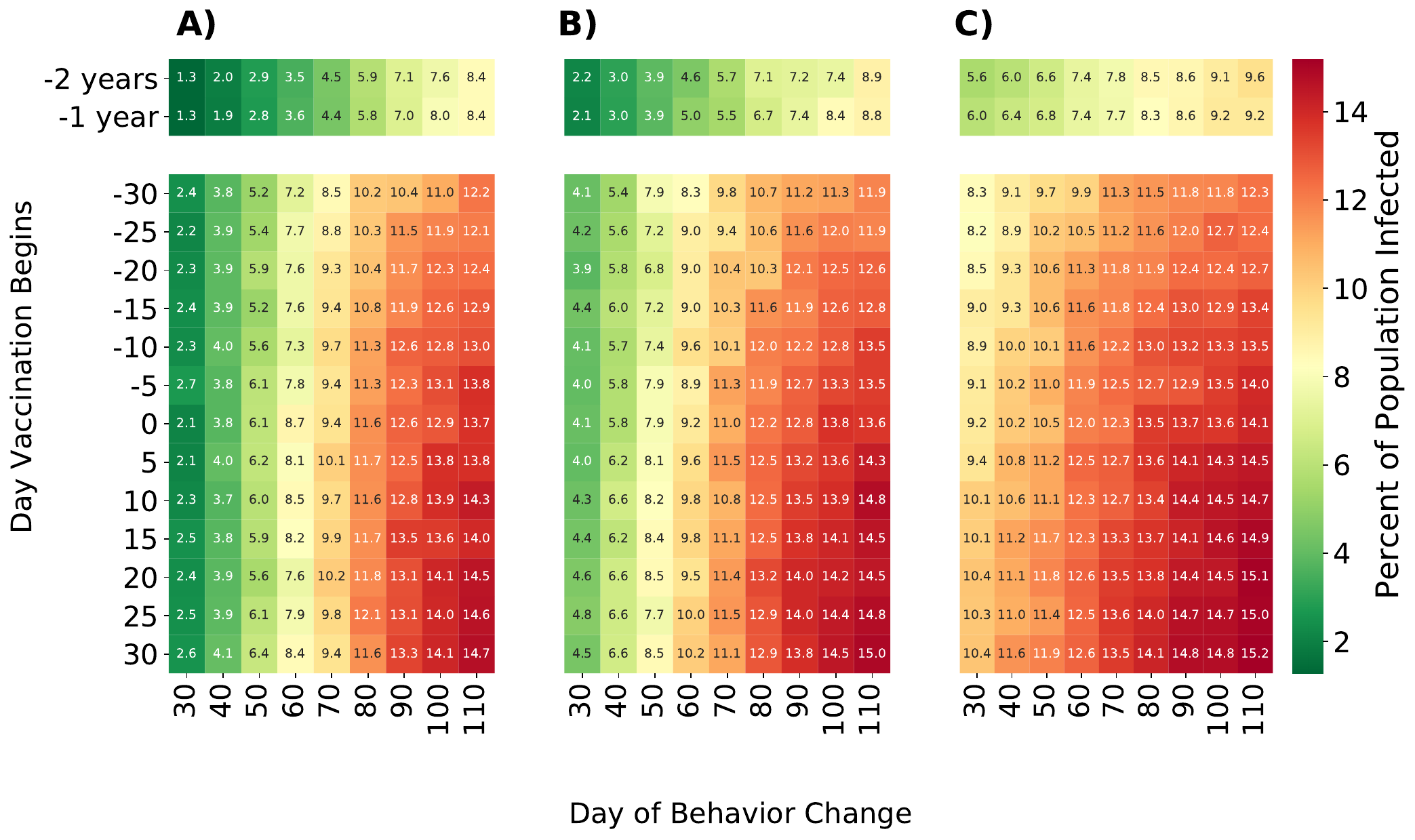}
\caption{\textbf{Percent of population infected with mpox after 250 days under different intervention timings and intensities.} Interventions only affect men in strata 5 and 6 of sexual activity. Cell values indicate the percent of the network infected after 250 days. Rows indicate the day that vaccines become available; negative numbers indicate vaccination becoming available prior to the start of the outbreak. Columns show simulations where individuals reduce their probability of having a one-time partner by 75\%, 50\%, and 25\%, respectively.}\label{heatmap}
\end{figure}

To better understand the impact of intervention timing, we compared the trade-off of the intensity of individual behavior change with the timing of the intervention. We compared three intensities of behavior change; reducing the probability of a one-time partner by 25\%, 50\%, or 75\%, and several different intervention timings. Vaccination initiation ranged from 30 days prior to any outbreak to 30 days after the outbreak began and behavior change initiation ranged from 30 to 110 days after the outbreak began (Figure \ref{heatmap}). As expected, earlier and more intense interventions result in fewer overall infections. Beginning vaccination a year before the outbreak results in only 5.5\% of men being infected, averting 950 infections or nearly 10\% of the total population in our model. However, it is evident that even for early initiation of a vaccination campaign, individual behavior modification has an important role, reducing the overall number of individuals infected from 15.2\% to 1.3\% of the network when comparing scenarios with the least intense and latest to start behavioral change to scenarios with the most intense and earliest behavior change.

\subsection{Infection Dynamics}
\begin{figure}[H]
\centering
\includegraphics[width = 1\textwidth]{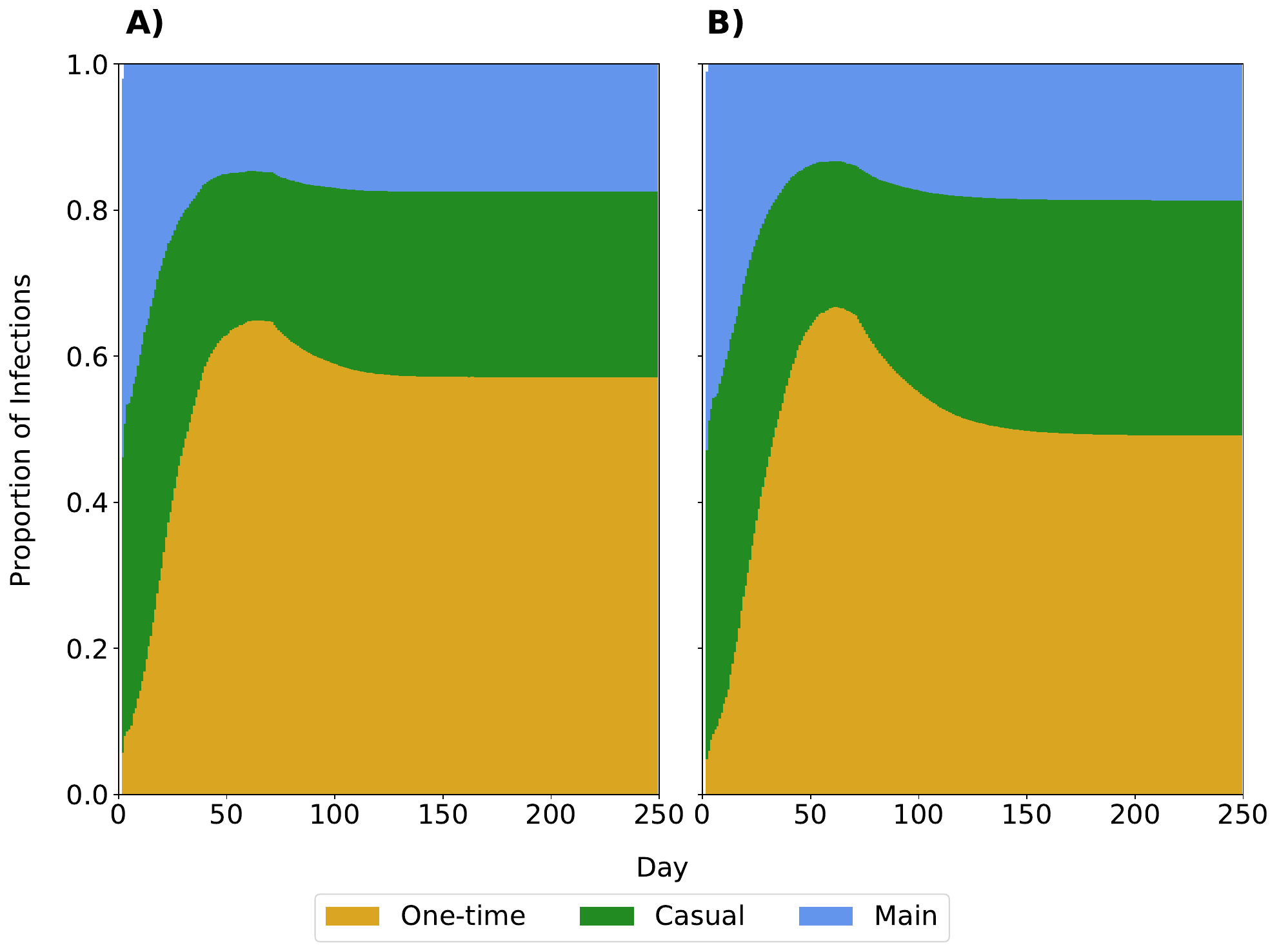}
\caption{\textbf{Proportion of infections attributable to each relationship type.} Panels indicate the proportion of cumulative infections by time $t$ attributable to each relationship type when intervention occurs only in the 25\% of men most likely to have a one-time partner. Panel A shows the results with full isolation compliance after diagnosis; Panel B shows results with partial compliance.}\label{inf_source}
\end{figure}

We tracked the source node of every infection in our model, allowing us to calculate the proportion of infections attributable to each relationship type. Figure \ref{inf_source} shows the proportion of infections at time $t$ attributable to each infection type with full isolation compliance (Panel A) and partial isolation compliance (Panel B) in the scenario with behavior change introduced on day 70 and vaccination at day 30. In either scenario, the early outbreak is driven by main and casual partnerships. Within 50 days, more than half of total infections are attributable to one-time partnerships. The relatively sharp decrease in the proportion of infections attributable to one-time partnerships is due to the introduction of behavior change on day 70, showcasing the impact of even 25\% of individuals reducing their probability of having a one-time partnership. In the partial isolation compliance scenario, when individuals are isolated, they still have some possibility of interacting with their main or casual partners. Thus, a larger proportion of infections come from these relationships. There is greater between-simulation variability in these proportions earlier, given that early infections may be driven by whether the initially infected nodes have main and casual relationships as well as the variability to the smaller number of total infections early in the outbreak (Figure \ref{inf_source_boxplot}).

\begin{figure}[H]
\centering
\includegraphics[width = 1\textwidth]{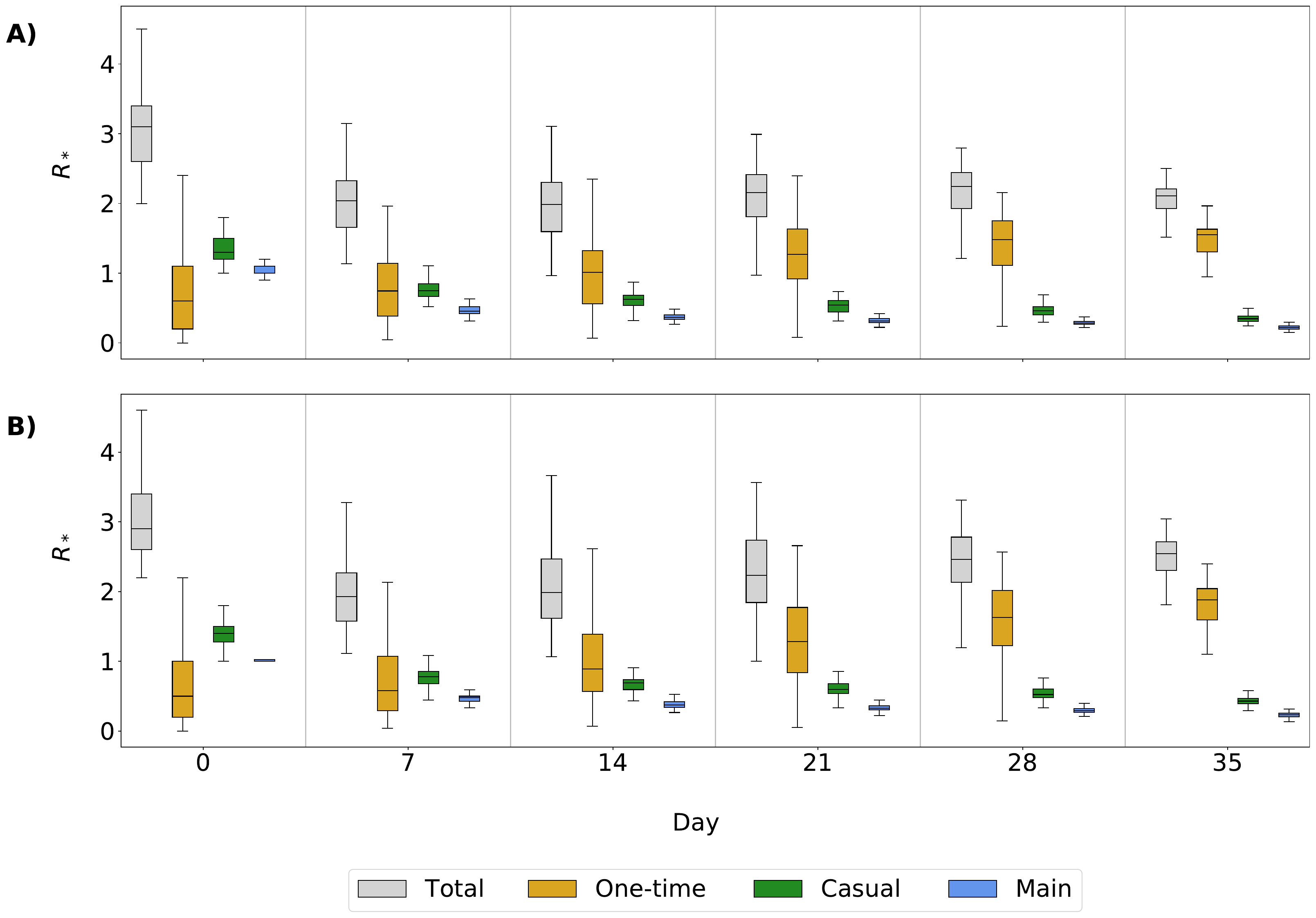}
\caption{\textbf{Effective reproductive number ($R^t_*$) at time $t$.} Box plots show the results of $R^t_*$ for main, casual, one-time, and total partnerships. Panel A shows results from simulations with full isolation compliance; Panel B shows results from simulations with partial isolation compliance. Time $t=0$ is a special case of $R^t_*$ which is equivalent to $R_0$.}\label{boxplot}
\end{figure}

This can also be seen in the $R^t_*$ values for each relationship type at different time steps (Figure \ref{boxplot}). Values at time $t=0$ are equivalent to $R_0$, the average number of infections per initially infected node. The $R^t_*$ value for one-time partnerships is initially lower than the $R^t_*$ values for main and casual partners but becomes relatively more important to overall transmission within the first 14 days of the simulation. At $t = 0$, median $R_*^t = $1.30 for casual partnerships ($P_{25\%}$ and $P_{75\%}$: 1.20, 1.50), 1.00 for main ($P_{25\%}$ and $P_{75\%}$: 1.00, 1.10), and 0.6 for one-time ($P_{25\%}$ and $P_{75\%}$: 0.20, 1.10). By $t = 28$, the median $R_*^t$ for one-time partnerships has more than doubled to 1.48 ($P_{25\%}$ and $P_{75\%}$: 1.11, 1.75). At the same time, it decreased for casual and main partnerships: median $R_*^t$ = 0.46 ($P_{25\%}$ and $P_{75\%}$: 0.40, 0.52) and 0.29 ($P_{25\%}$ and $P_{75\%}$: 0.27, 0.31), respectively. This change in the $R_*^t$ values indicates that the continuation of the outbreak after this point is largely due to spread via one-time partnerships.

\section{Discussion}\label{sec12}

This study integrated data about the clinical course of mpox, sexual network characteristics of MSM, and individual- and policy-level interventions to generate a mechanistic network model of mpox transmission in a US-based population of MSM. We describe simulations that compare the impact of the timing of interventions at different levels of compliance and intervention access. 

We found that individual behavior change, reducing the probability of having a one-time partner, can reduce the number of individuals infected, even if implemented only among individuals with high-risk sexual behavior.The timing of interventions also played a crucial role in the final outcome of our modeling. We show that earlier implementation of behavior change and vaccination has a greater impact on reducing the spread of mpox. This is particularly true if behavior change begins in the early phases of an outbreak when vaccine coverage may be limited, suggesting that rapid deployment of public health messaging and vaccination campaigns at the onset of an outbreak is vital. However, even when vaccination was delayed, individual behavior modification mitigated transmission. This finding is particularly relevant in settings where vaccine supply may be limited or slow to roll out.

While intense and early behavior change can clearly reduce mpox transmission in our simulations, it can be burdensome and difficult to sustain. We present different combinations of intervention intensity and timing for behavior change with vaccination timing. Our results suggest that better epidemic preparedness, namely better vaccine coverage prior to the start of the epidemic, can offset variations in individual behavior change. Even in scenarios where only high-risk individuals were vaccinated, our study showed that beginning vaccination a year prior to the outbreak can be as effective at preventing transmission as behavior change. If intervening on a subset of the population with the highest risk leads to a sufficient decrease in cumulative infection, it could be possible to tailor public messaging and communication campaigns to this group.

Finally, our model contributes insight into the differential impact of relationship types on transmission dynamics over time. By tracking the source of each infection during modeling, we are able to present $R_0$ and $R_*^t$ values for each relationship type and at different times during the outbreak. Our results show that main and casual partnerships drive the early spread of infections, while one-time relationships connect the sparse network over time. These connections prevent the infection from remaining constrained to a limited portion of the network, particularly in scenarios where behavior change is delayed or limited. Additionally, we show that individuals who are most likely to form one-time partnerships have, on average, a greater number of at-risk (serodiscordant) sexual interactions than individuals with the same average number of partners but whose partners are sustained. These results are particular to the mpox outbreak in 2022, but could inform responses to future mpox outbreaks, or STIs more broadly, by underlining the importance of widespread vaccination as well as the impact of timely behavioral adaptations during STI outbreaks.

Our results showing the protective effect of behavior change and vaccination during the 2022 outbreak are consistent with those of studies using other modeling approaches, such as statistical network models\cite{Clay2024ModellingOutbreak, Spicknall2022Morbidity2022, VanDijck2023TheStudy}. We found a greater impact of behavior change alone than Clay et al., approximately 30\% of cases were averted in our scenario after 250 days compared to 25\% after 1 year. However, they found a greater impact of vaccination and behavior change together, averting 84\% of cases after a year compared to 30\% after 250 days in our results. The greater impact of vaccination in the modeling from Clay et al. could be due to the longer time frame, given that vaccine roll out was relatively slow. Additional sources of variation could be due to different assumptions about the level of behavioral change, vaccine efficacy, and the natural history of the disease. 

Comparing disparate methods gives us greater confidence in common results, such as the efficacy of behavioral adaptation and vaccination in outbreak prevention. However, the mechanistic modeling approach allows us unique insight into this outbreak and potential future outbreaks of mpox that other modeling approaches cannot capture. First, we are able to incorporate individual-level action, such as heterogeneity in sexual behavior and changes in partnership seeking. We are also able to capture more realistic sexual behavior that other popular modeling approaches, such as compartmental models, are unable to model, namely, the importance of repeated sexual partnerships. Secondly, our model is able to track the source of each infection, so we are able to show the proportion of infections attributable to each relationship type in each simulation scenario, providing important information about the effect of behavior change and isolation on the dynamics of the outbreak. Finally, our model is readily interpretable and flexible. By modeling and intervening on the network at the level of the individual, translating modeling results to potential disease prevention recommendations, or updating the model with new information and exploring new interventions, is relatively straightforward.

Our study has several limitations. First, while this model parameterized the simulated sexual networks based on survey data from MSM in the US, this may not be generalizable to other populations of MSM\cite{Weiss2020EgocentricStudy, Jenness2016ImpactFor}, and is not directly generalizable to a heterosexual population. The network parameter estimates we use are pulled from multiple studies, which could introduce additional uncertainty in the model. However, this approach is also taken in other modeling studies for mpox \cite{Clay2024ModellingOutbreak, Spicknall2022Morbidity2022} and has the benefit of allowing our results to be more easily compared because the network parameterization is more similar. Secondly, the assumptions about isolation and diagnosis delays were based on early outbreak data, which may not fully reflect evolving clinical practices and public awareness as the outbreak progressed. Future studies could explore the impact of varying these parameters to reflect different healthcare settings and policy environments that impact healthcare access. Future work could also extend this modeling strategy to include assortative partner selection by individuals' demographics, such as age and race or ethnicity, and the potential impact of co-infection with other STIs, particularly HIV. Finally, we are limited by the lack of computationally scalable inferential techniques for mechanistic network models. Despite these limitations, our modeling provides valuable and unique insights.

\section{Conclusions}\label{sec13}

Mechanistic network models provide a valuable framework for understanding the transmission dynamics of mpox in MSM populations. Our results suggest that early and sustained public health interventions, combining behavior change with vaccination, are critical to controlling the spread of mpox. Tailoring interventions to target high-risk groups, particularly those engaging in one-time partnerships, can optimize the use of limited resources and achieve meaningful reductions in transmission. These insights have important implications for public health strategies in the face of future outbreaks of mpox or similar STIs.

\section{Abbreviations}

\textbf{CDC}: Centers for Disease Control and Prevention \\
\textbf{HIV/AIDS}: Human immunodeficiency virus/Acquired immunodeficiency syndrome \\
\textbf{MSM}: Gay, bisexual, and other men who have sex with men \\
\textbf{STI}: Sexually transmitted infection \\

\backmatter


\bmhead*{Declarations}

\bmhead*{Ethics approval and consent to participate}

Not applicable.

\bmhead*{Consent for publication}

Not applicable.

\bmhead*{Availability of data and materials}

All data and code used for this manuscript have been made publicly available at \url{https://github.com/onnela-lab/mpox-model}.

\bmhead*{Competing interests}

The authors declare that they have no competing interests.

\bmhead*{Funding}

This project was supported by the NIH grants R01AI138901 and T32AI007358. The funder had no role in the study design, collection, analysis, or interpretation of the data, writing the manuscript, or the decision to submit the paper for publication.

\bmhead*{Authors' contributions}

E.G.C. and J.P.O. designed the research; E.G.C. performed the research. E.G.C. and J.P.O. wrote and edited the paper. J.P.O. supervised the research.

\bmhead*{Acknowledgements}

We thank the members of the Onnela lab for their insights and feedback on this project.

\bmhead*{Authors' information}

Department of Biostatistics, Harvard TH Chan School of Public Health, Boston, MA.

\begin{appendices}

\section{Supplementary Information}\label{secA1}

\subsection{Vaccination Information}
\begin{figure}[H]
    \centering
    \includegraphics[width=0.7\linewidth]{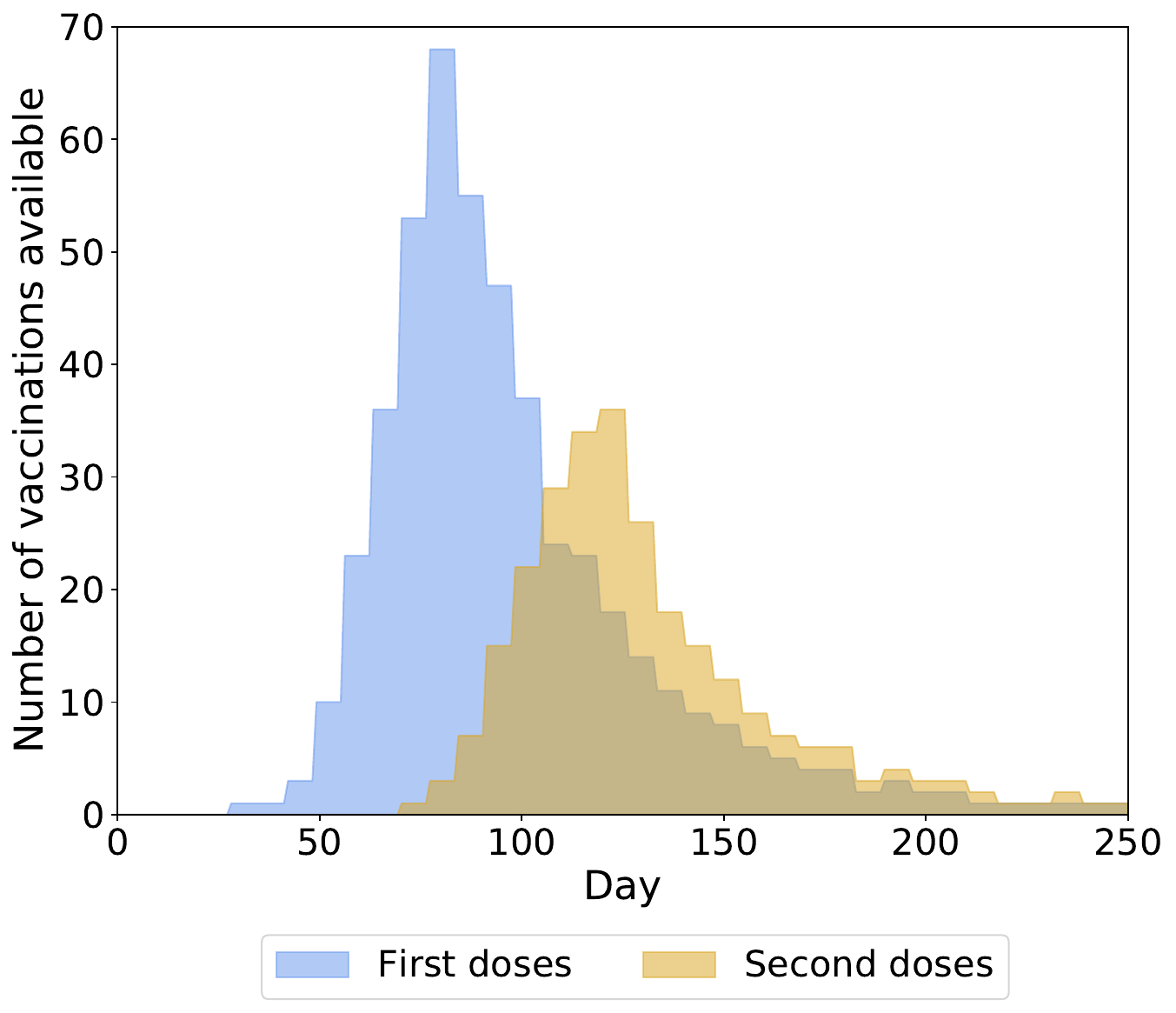}
    \caption{\textbf{Vaccine availability by day and dose type.}}
    \label{vax-avail}
\end{figure}

The number of vaccines administered in the modeling reflects the number of first and second doses of JYNNEOS vaccines received in the US from May 22, 2022 to July 1, 2023. The raw data was obtained on July 11, 2023 from the CDC's website\cite{CDCMpoxU.S.}. To account for the fact that not everyone who received a vaccine was male, the number of first doses given was multiplied by 0.91, as 91\% of those who received a first dose were male, and the number of second doses was multiplied by 0.94 for the same reason, per the CDC's information on vaccine administration. To account for population, we divided the number of doses by the total population at risk, 1,998,039\cite{CDCJYNNEOSJurisdiction, Owens2023JYNNEOS2023}. Therefore, we were able to approximate the number of vaccines available in a given week in a population of 10,000 individuals per the national averages. Daily numbers were obtained by dividing the weekly numbers by 7. Figure \ref{vax-avail} shows the number of vaccines available in our simulation per day by dose type.

\subsection{Sensitivity Analyses}

\subsubsection{Isolation Compliance}

\begin{figure}[H]
\centering
\includegraphics[width = 1\textwidth]{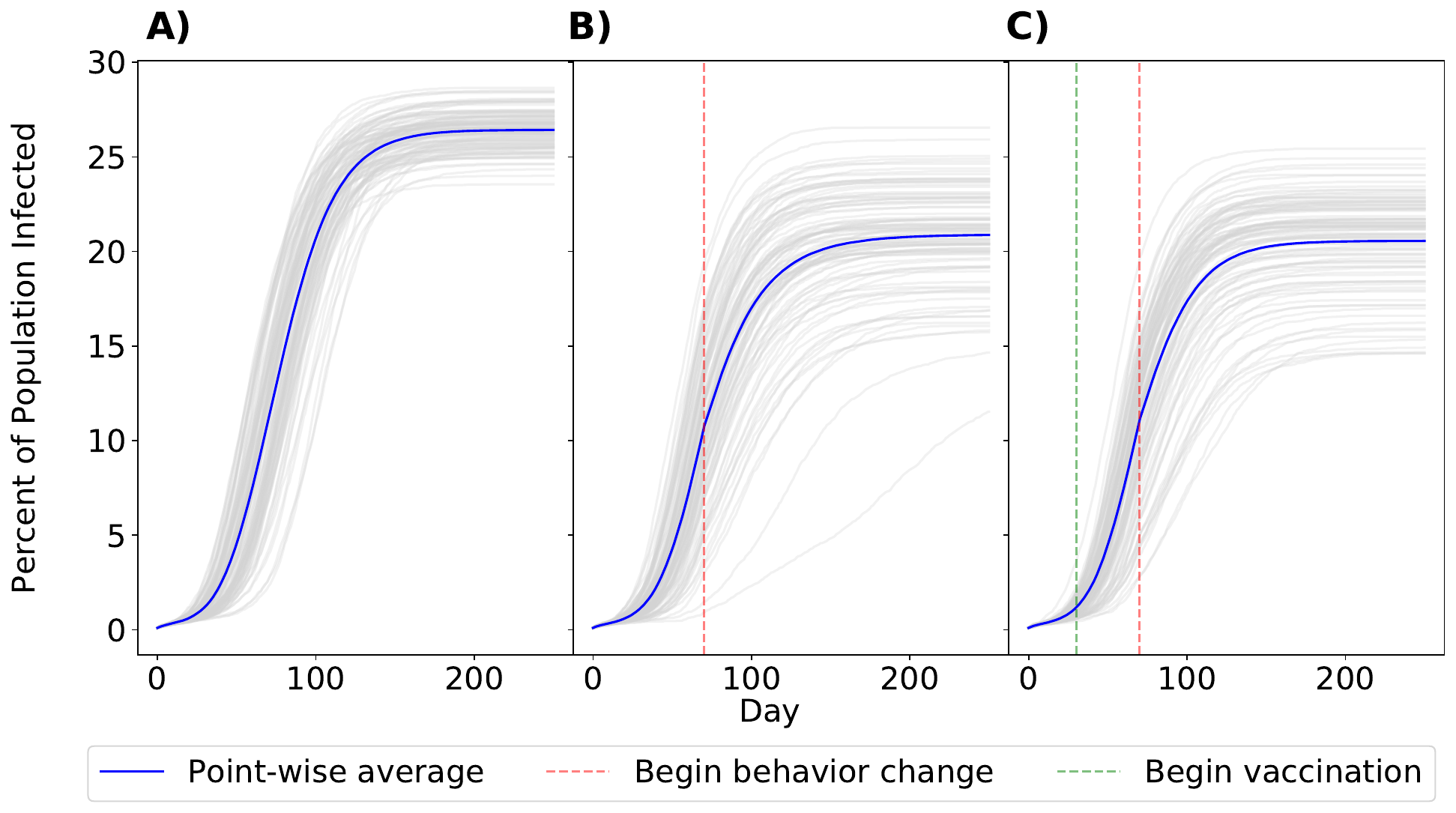}
\caption{\textbf{Comparison of universal interventions with partial isolation compliance.} Panels indicate the percent of network infected with mpox after 250 days with no intervention (Panel A), universal behavior change (Panel B), or universal behavior change with vaccination (Panel C). Grey lines denote individual simulations. The point-wise average is shown in blue. Vertical lines indicate the day of intervention initiation.}\label{sens1}
\end{figure}

As a sensitivity analysis, we also present the results comparing the percent of the population that becomes infected after 250 days if we assume a less optimistic isolation scheme in which, once diagnosed, individuals only fully isolate from one-time partnerships but partially isolate from main and casual partnerships. Comparing Figure \ref{sens1} to the main results in Figure \ref{compint}, partial compliance leads to a greater number of overall infections but does not change the conclusion that behavioral change and vaccination reduce the final size of the outbreak. Without intervention and with only partial compliance the 26.43\% of the population is infected after 250 days, ($P_{25\%}$ and $P_{75\%}$: 25.67\%, 27.15\%). With universal intervention, this decreases to 20.56\% ($P_{25\%}$ and $P_{75\%}$: 19.19\%, 22.34\%). 

\begin{figure}[H]
\centering
\includegraphics[width = 1\textwidth]{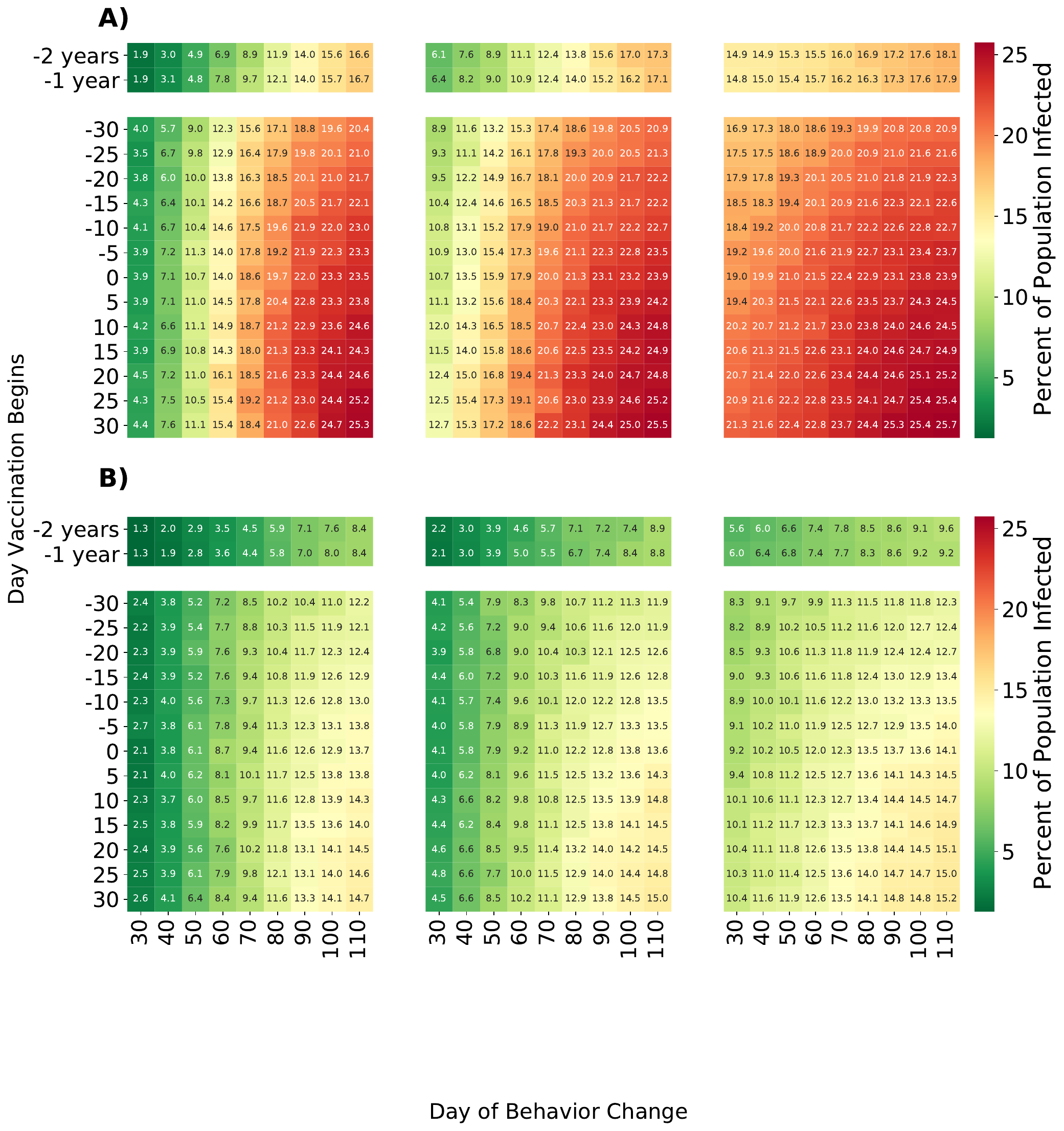}
\caption{\textbf{Percent of the population infected with mpox after 250 days under different intervention timings and intensities and with different levels of isolation compliance.} Interventions only affect men in strata 5 and 6 of sexual activity. Panel A shows results from simulations with partial isolation compliance; Panel B shows results from simulations with full isolation compliance. Cell values indicate the percent of the network infected after 250 days. Rows indicate the day that vaccines become available; negative numbers indicate vaccination becoming available prior to the start of the outbreak. The left, middle, and right columns show simulations where individuals reduce their probability of having a one-time partner by 75\%, 50\%, and 25\%, respectively.}\label{heatmap_both}
\end{figure}

Figure \ref{heatmap_both} shows the results at different intervention timings and intensities with partial isolation compliance (Panel A) and full compliance (Panel B). Panel B contains the same information as Figure \ref{heatmap}, but the colors are recalculated to include the entire range in Panel A. We can see that greater isolation compliance (Panel B) can be as protective at the population level as more intense and earlier intervention (Figure \ref{heatmap_both}). For example, full isolation compliance with vaccination beginning on day 30 and 50\% behavior change beginning on day 70 (11.1\% of the population infected) is similar to partial isolation compliance, but with more intense behavior change and both interventions beginning 20 days earlier.

\begin{figure}[H]
\centering
\includegraphics[width = 1\textwidth]{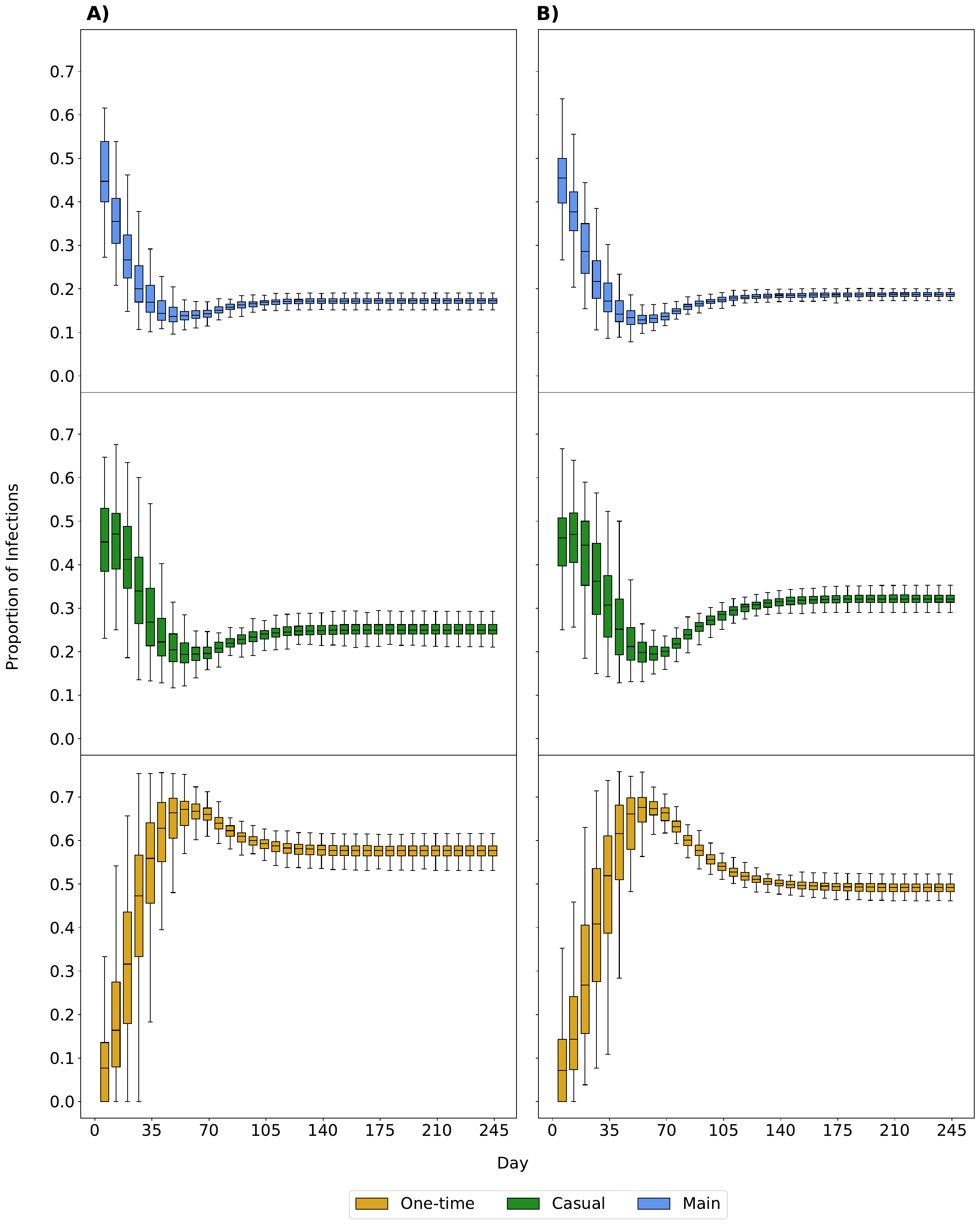}
\caption{\textbf{Proportion of infections attributable to each relationship type.} Panels indicate the proportion of cumulative infections by time $t$ attributable to each relationship type when intervention occurs in the 25\% of men most likely to have a one-time partner. Panel A shows the results with full isolation compliance after diagnosis; Panel B shows partial compliance. Box plots show results of 100 independent simulations are presented for every 7 days of data.}\label{inf_source_boxplot}
\end{figure}

\subsubsection{Infection Parameters}

\begin{figure}[H]
\centering
\includegraphics[width = 1\textwidth]{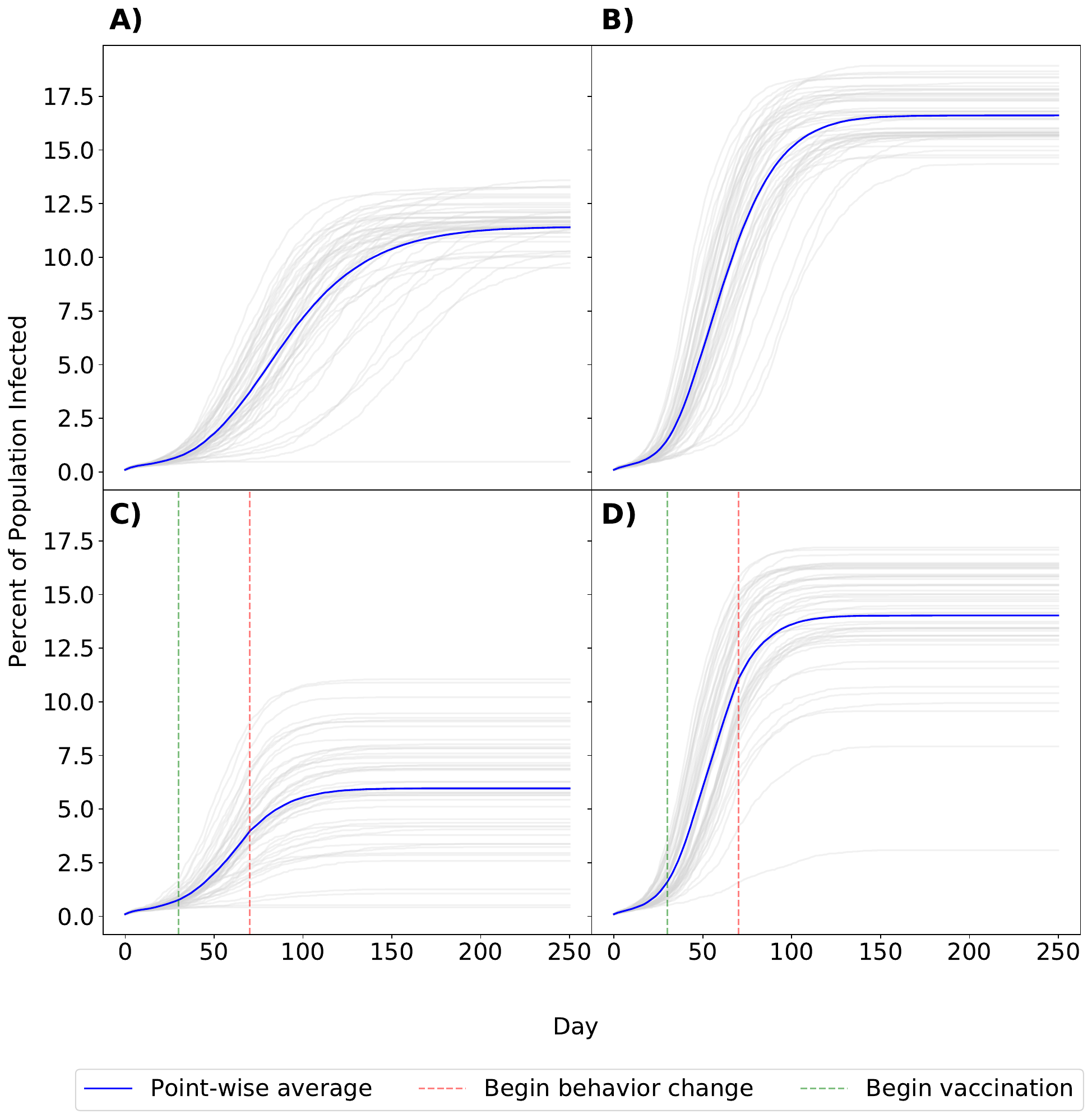}
\caption{\textbf{Comparison of Infection Parameters.} Panels indicate the percent of network infected with mpox after 250 days with no intervention and optimistic infection parameters (Panel A), no intervention and pessimistic infection parameters (Panel B), universal behavior change with vaccination and optimistic infection parameters (Panel C), or universal behavior change with vaccination and pessimistic infection parameters (Panel C). Grey lines denote 50 individual simulations. The point-wise average is shown in blue. Vertical lines indicate the day of intervention initiation.}\label{bestworst}
\end{figure}

Due to the uncertainty around the clinical features of mpox, particularly the length of time individuals are in the exposed and infectious states; we conducted additional sensitivity analyses to ensure our findings are robust to the range of current estimates of the infection parameters. We selected incubation times from a review of published infection-to-onset times, which corrected for right-truncation bias\cite{Ponce2024IncubationEstimates}, and for the length of time individuals are infectious, we use the standard 2-4 week range. We created two scenarios: one which used the estimates that were most likely to lead to few infections that had a long incubation period and a short infection period (5.6 days and 2 weeks, respectively), the optimistic scenario, and one which used estimates most likely to lead to many infections that had a short incubation period and a long infection period (9.9 days and 4 weeks, respectively), the pessimistic scenario. Figure \ref{bestworst} shows the results of these scenarios, with the same interventions as the main results (reduction of one-time partners at 70 days and vaccination at 30 days). As expected, we see that the optimistic scenario leads to fewer total infections than our main results (mean percent of population infected without intervention: 11.40\%, 25th and 75th percentiles of infections:  11.30\%, 12.11\%; mean with intervention: 5.96\%, 25th and 75th percentiles of infections: 4.18\%, 7.77\%), while the worst-case scenario yields more infections (mean percent of population infected without intervention: 16.61\%, 25th and 75th percentiles of infections: 15.70\%, 17.57\%; mean with intervention: 14.02\%, 25th and 75th percentiles of infections: 13.08\%, 15.87\%). However, in both scenarios, we do see that the interventions reduce the number of overall infections.

\begin{figure}[H]
\centering
\includegraphics[width = 1\textwidth]{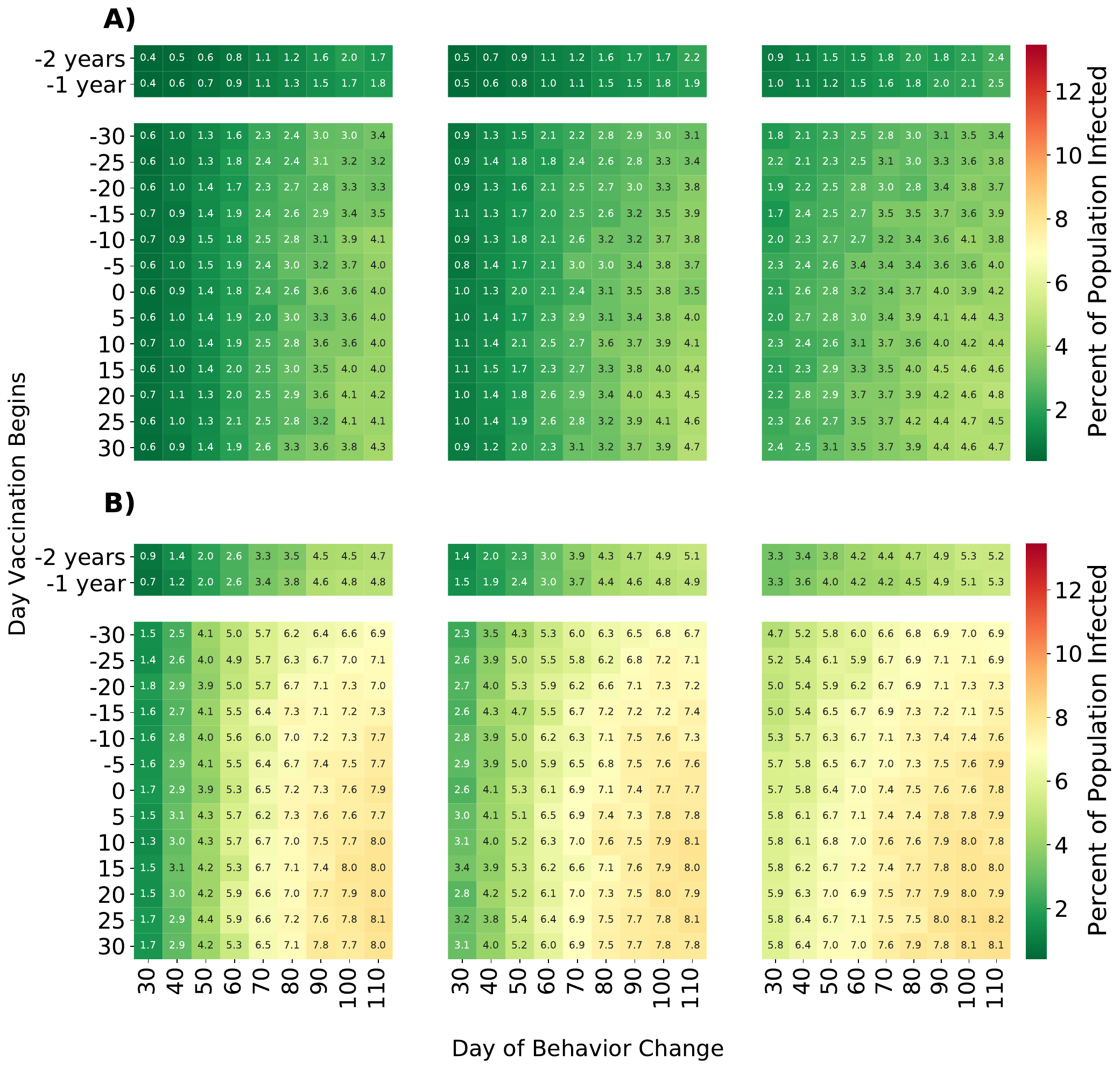}
\caption{\textbf{Percent of the population infected with mpox after 250 days under different intervention timings and intensities and with optimistic and pessimistic infection parameters with full isolation compliance and high transmission probability.} Interventions only affect men in strata 5 and 6 of sexual activity. Probability of transmission for sexual contact in a serodiscordant pair is 0.9. Panel A shows results from simulations with optimistic infection parameters; Panel B shows results from simulations with pessimistic infection parameters. Cell values indicate the percent of the network infected after 250 days. Rows indicate the day that vaccines become available; negative numbers indicate vaccination becoming available prior to the start of the outbreak. The left, middle, and right columns show simulations where individuals reduce their probability of having a one-time partner by 75\%, 50\%, and 25\%, respectively.}\label{heatmap_isol1_BW_highprob}
\end{figure}

\begin{figure}[H]
\centering
\includegraphics[width = 1\textwidth]{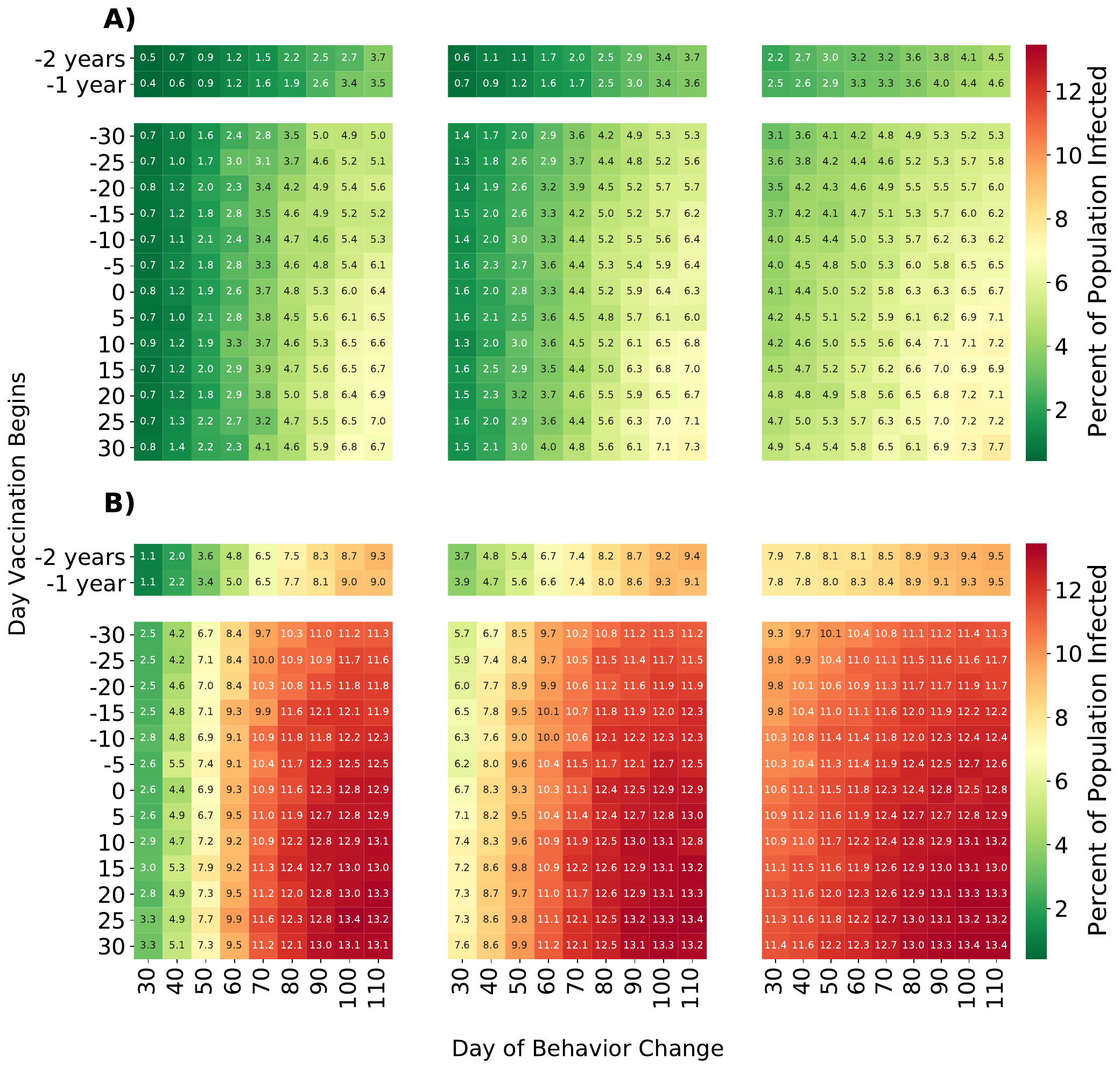}
\caption{\textbf{Percent of the population infected with mpox after 250 days under different intervention timings and intensities and with optimistic and pessimistic infection parameters with partial isolation compliance and high transmission probability.} Interventions only affect men in strata 5 and 6 of sexual activity. Probability of transmission for sexual contact in a serodiscordant pair is 0.9. Panel A shows results from simulations with optimistic infection parameters; Panel B shows results from simulations with pessimistic infection parameters. Cell values indicate the percent of the network infected after 250 days. Rows indicate the day that vaccines become available; negative numbers indicate vaccination becoming available prior to the start of the outbreak. The left, middle, and right columns show simulations where individuals reduce their probability of having a one-time partner by 75\%, 50\%, and 25\%, respectively.}\label{heatmap_isol2_BW_highprob}
\end{figure}

\begin{figure}[H]
\centering
\includegraphics[width = 1\textwidth]{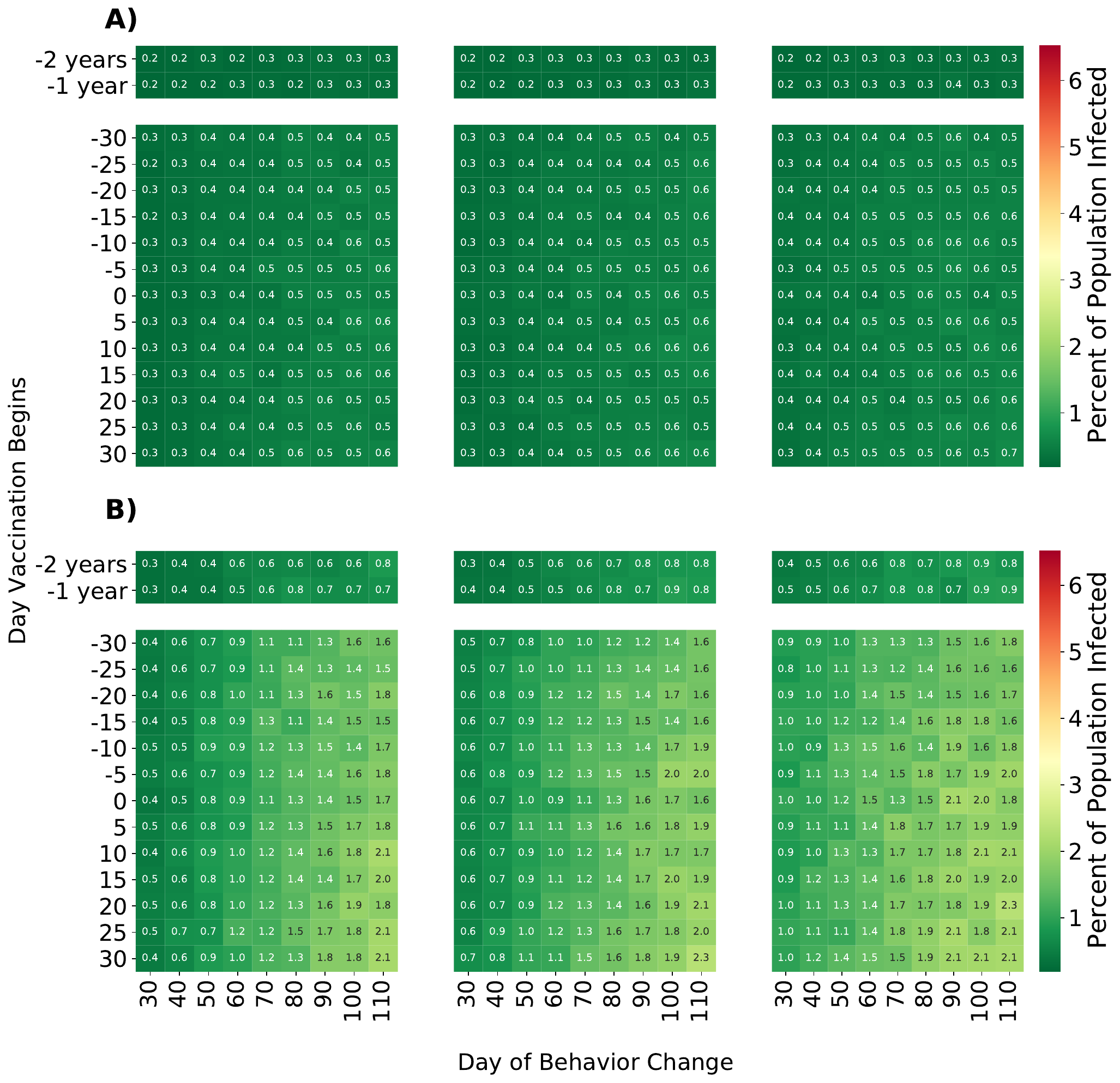}
\caption{\textbf{Percent of the population infected with mpox after 250 days under different intervention timings and intensities and with optimistic and pessimistic infection parameters with full isolation compliance and low transmission probability.} Interventions only affect men in strata 5 and 6 of sexual activity. Probability of transmission for sexual contact in a serodiscordant pair is 0.5. Panel A shows results from simulations with optimistic infection parameters; Panel B shows results from simulations with pessimistic infection parameters. Cell values indicate the percent of the network infected after 250 days. Rows indicate the day that vaccines become available; negative numbers indicate vaccination becoming available prior to the start of the outbreak. The left, middle, and right columns show simulations where individuals reduce their probability of having a one-time partner by 75\%, 50\%, and 25\%, respectively.}\label{heatmap_isol1_BW_lowprob}
\end{figure}

\begin{figure}[H]
\centering
\includegraphics[width = 1\textwidth]{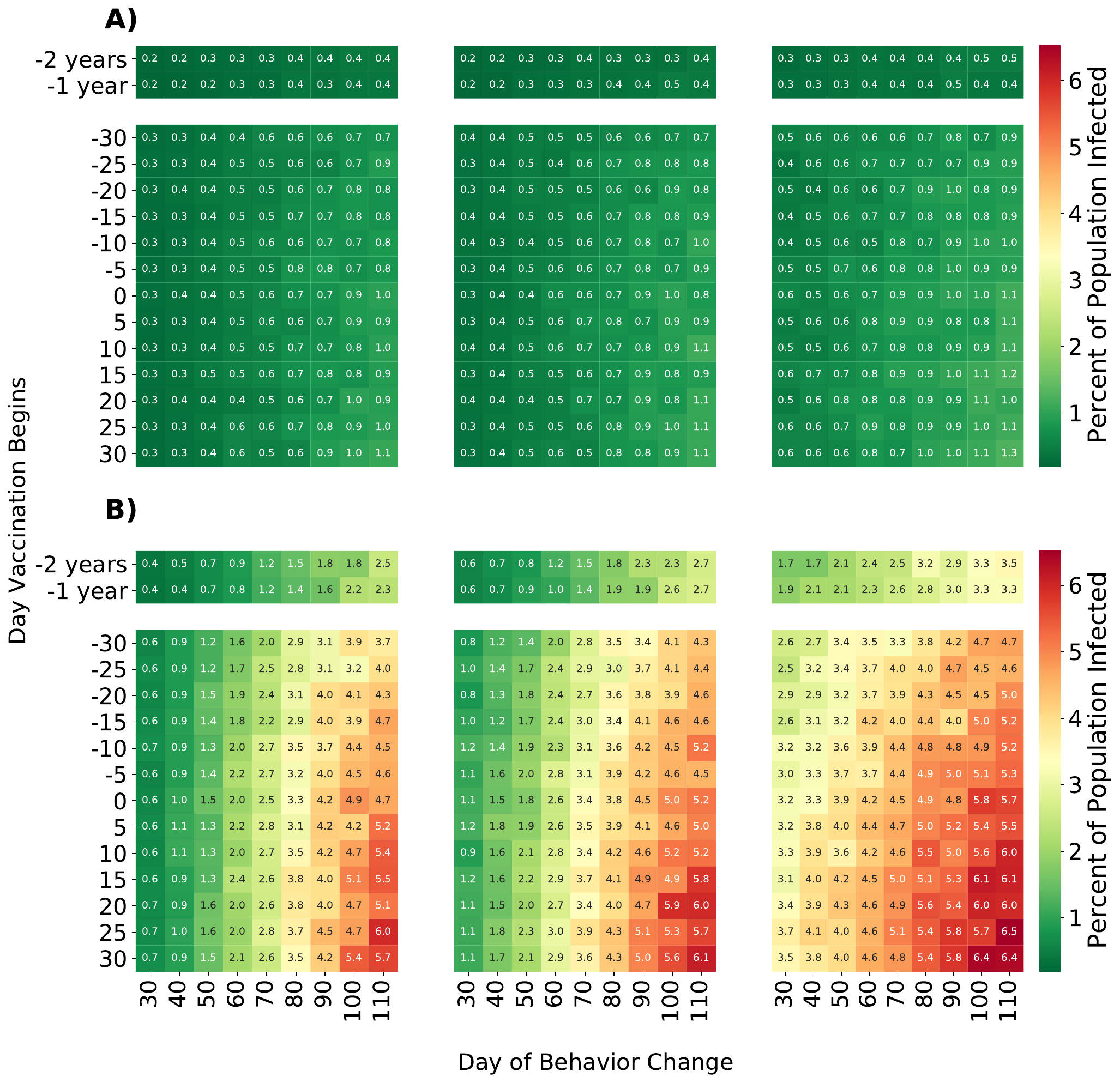}
\caption{\textbf{Percent of the population infected with mpox after 250 days under different intervention timings and intensities and with optimistic and pessimistic infection parameters with partial isolation compliance and high transmission probability.} Interventions only affect men in strata 5 and 6 of sexual activity. Probability of transmission for sexual contact in a serodiscordant pair is 0.5. Panel A shows results from simulations with optimistic infection parameters; Panel B shows results from simulations with pessimistic infection parameters. Cell values indicate the percent of the network infected after 250 days. Rows indicate the day that vaccines become available; negative numbers indicate vaccination becoming available prior to the start of the outbreak. The left, middle, and right columns show simulations where individuals reduce their probability of having a one-time partner by 75\%, 50\%, and 25\%, respectively.}\label{heatmap_isol2_BW_lowprob}
\end{figure}

Figures \ref{heatmap_isol1_BW_highprob}, \ref{heatmap_isol2_BW_highprob}, \ref{heatmap_isol1_BW_lowprob}, and \ref{heatmap_isol2_BW_lowprob} are joint sensitivity analyses showing the results of varying the infection parameters (best and worst case scenarios presented above), the probability of transmission during sexual contact, and isolation scenarios over a range of intervention timings. These joint sensitivity analyses demonstrate that the only scenario in which the interventions do not impact cumulative infections is in simulations under the most optimistic combination of infection parameters and a low probability of transmission during sexual contact due to lack of overall transmission of mpox in the population (Figure \ref{heatmap_isol1_BW_lowprob} panel A). , The consistent conclusions from our sensitivity analyses indicates that our results are robust even if the infection parameters vary.

\subsubsection{Population Size}
\begin{figure}[H]
\centering
\includegraphics[width = 1\textwidth]{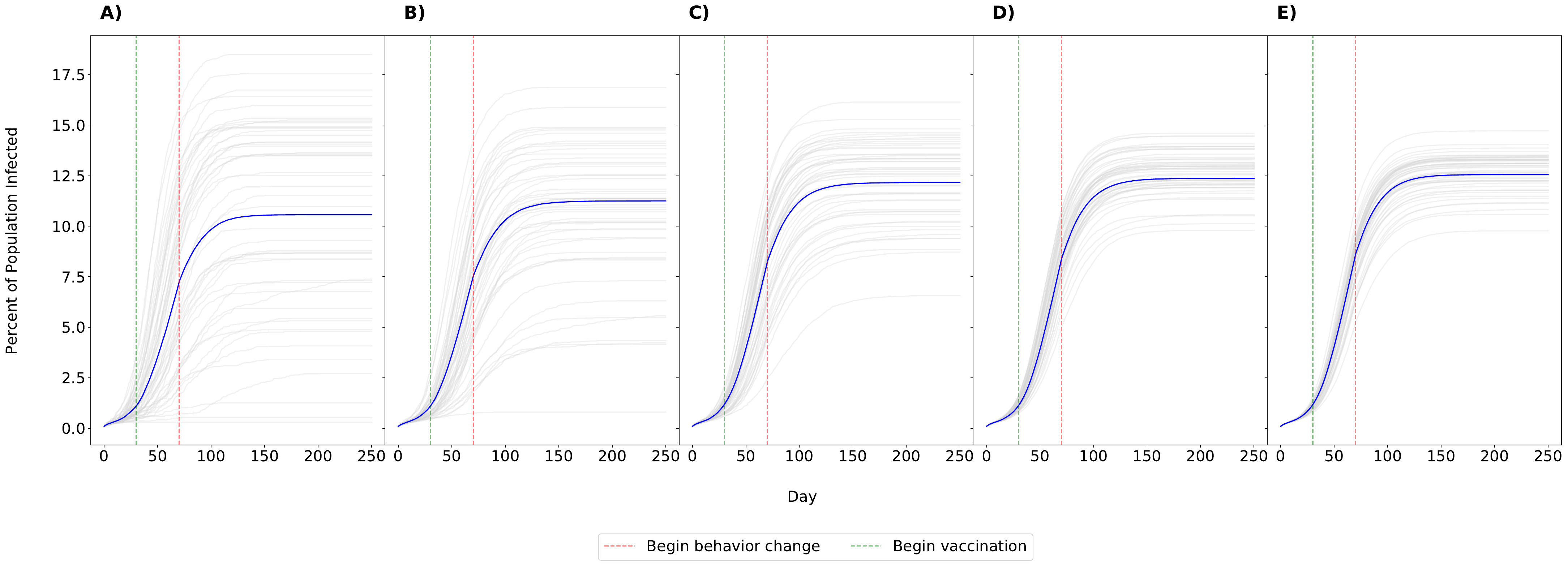}
\caption{\textbf{Cumulative incidence with different population sizes.} Panels indicate the percent of network infected with mpox after 250 days in a population of 5,000 nodes (A), 10,000 nodes (B), 20,000 nodes (C), 40,000 nodes (D), or 80,000 nodes (E). Grey lines denote individual simulations. The point-wise average is shown in blue. Vertical lines indicate the day of intervention initiation. The figure shows the runtime of 50 independent simulations the main intervention scenario with behavior change and vaccination.}\label{popsize}
\end{figure}

\begin{figure}[H]
\centering
\includegraphics[width = 1\textwidth]{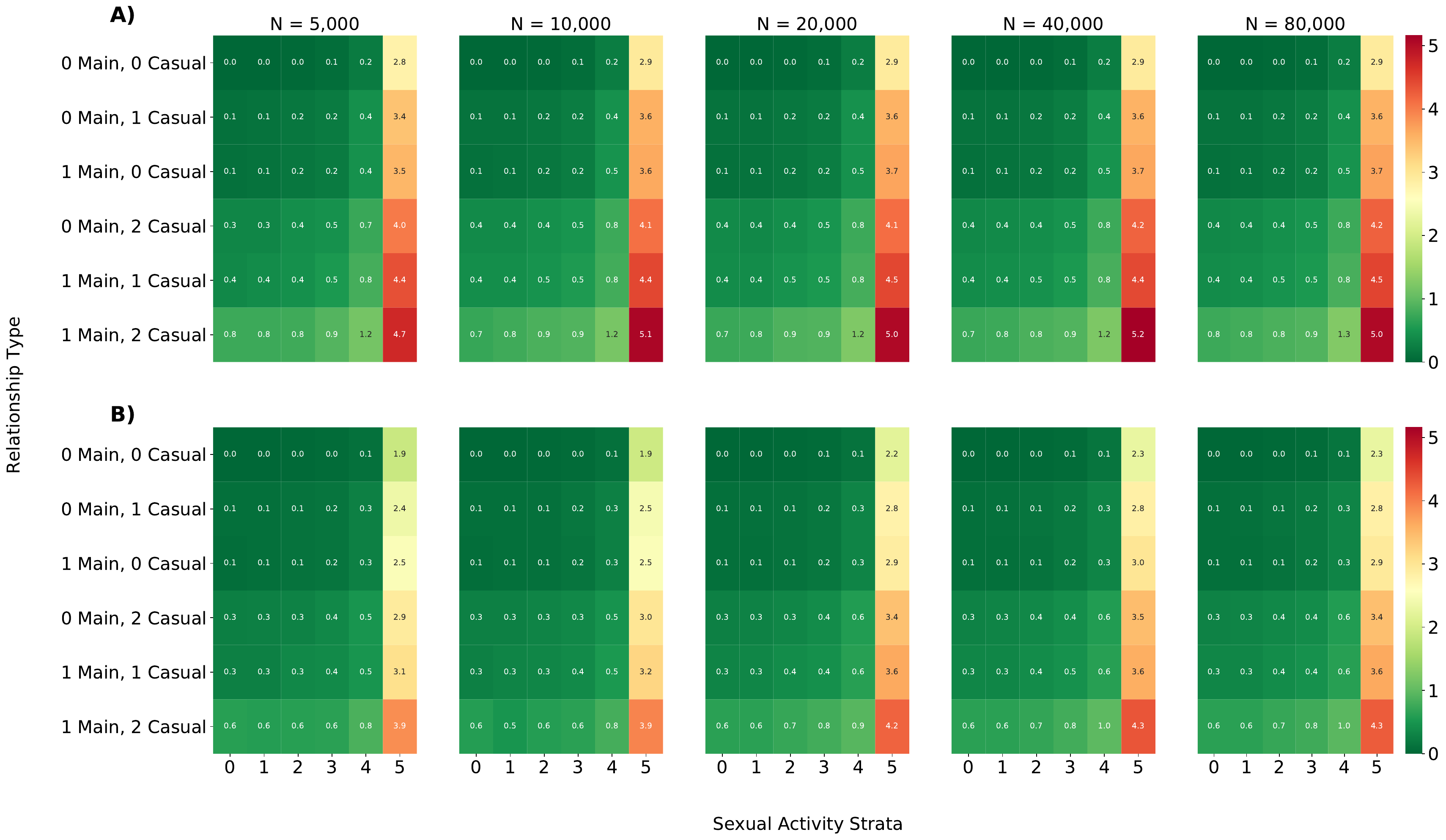}
\caption{\textbf{Comparison of serodiscordant sexual interactions by relationship type and sexual activity strata} Cells indicate the average number of serodiscordant sexual interactions individuals of a given relationship type and sexual activity stratum have over 250 days with no intervention (Panel A) or intervention only in the 25\% of men most likely to have a one-time partner (Panel B). Rows indicate relationship type (preferred number of main and casual partners), while columns indicate sexual activity strata, or a node's daily probability of having a one-time partner}\label{contacts_popsize}
\end{figure}

To better understand the sensitivity of our results to the total population size in the simulation, we repeated the simulations for varying population sizes. Figure \ref{popsize} demonstrates the results of 50 simulations for networks with N = 5,0000, N = 10,000, N = 20,000, N = 40,000, and N = 80,000 nodes. The mean percent of the network infected after 250 days is consistent for different network sizes, ranging from 10.57\% for the network of 5,000 nodes to 12.6\% for the network of 80,000 nodes. However, the network size has a large effect on the between-simulation variability: The 25th and 75th percentiles of infections in the network of 5,000 nodes are 2.62\% and 13.58\%, whereas for the network of 80,000 nodes the 25th and 75th percentiles of infections are 6.56\% and 12.94\%. In particular, the smaller networks are more likely to see the epidemic end early by chance (final infection percentage close to 0). Similarly, population size also does not seem to greatly affect the number of at-risk sexual contacts individuals have on average. In Figure \ref{contacts_popsize}, we can see that the variation in number of at-risk contacts is much greater between relationship types and sexual activity strata than between network population sizes.

\begin{figure}[H]
\centering
\includegraphics[width = 1\textwidth]{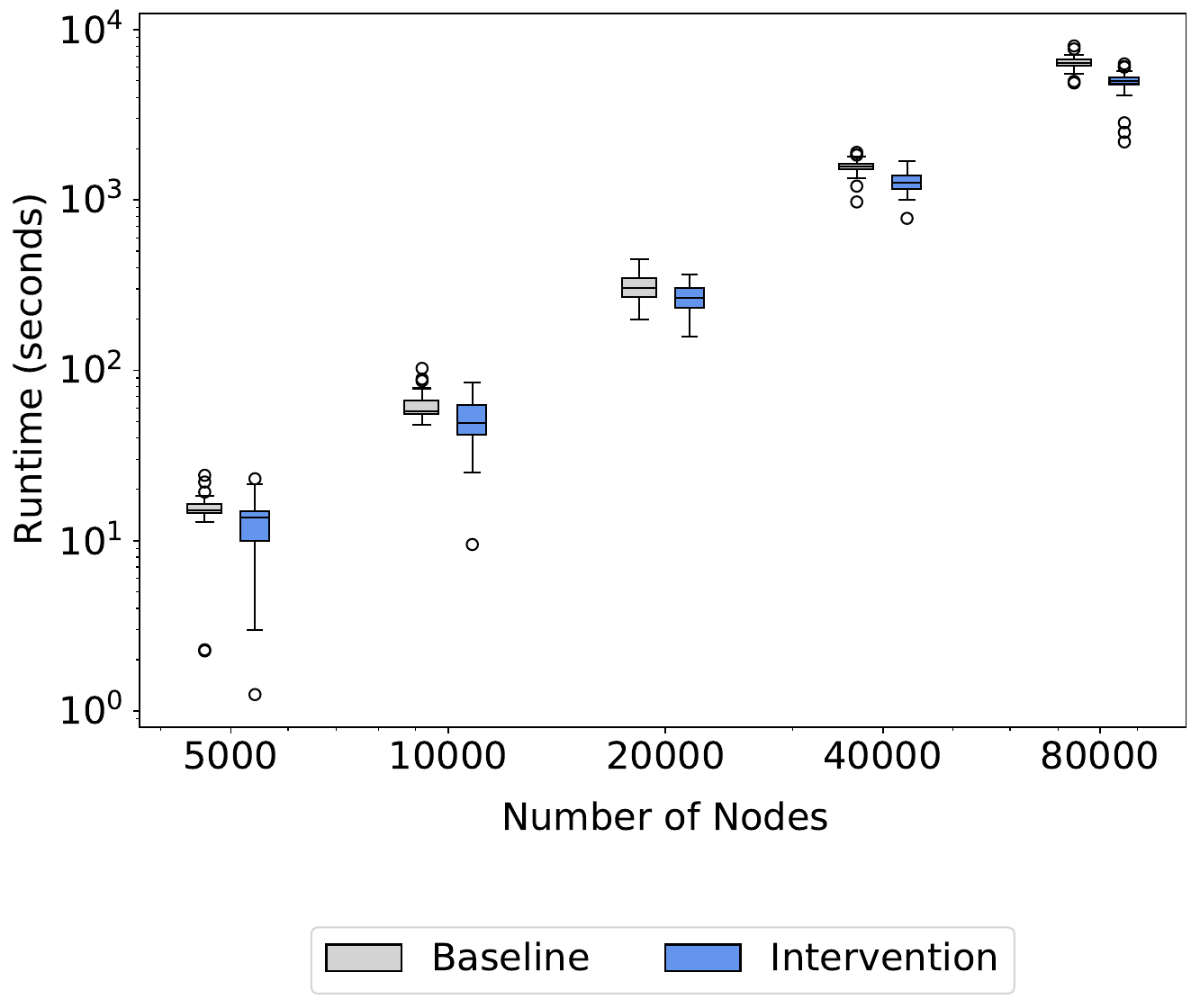}
\caption{\textbf{Comparison of Runtime for Network Populations} The figure shows the runtime of 50 independent simulations of the baseline, no-intervention scenario as well as as the main intervention scenario with behavior change and vaccination. All simulations are run for 250 time steps.}\label{runtime}
\end{figure}

We also calculated the effect of network size on the runtime of the algorithm. Figure \ref{runtime} shows the runtime in seconds of 50 simulations at different network population sizes. As network size doubles, the runtime increases approximately 4-fold, yielding a $O(n^2)$ computational complexity. Additionally, algorithmic complexity does not differ greatly between running the baseline (no intervention) scenario or simulations with the behavior change and vaccination interventions.

\subsubsection{Network Dynamics and Structure}
Due to the dynamic nature of the network, edge rewiring, or partner re-pairing, must be performed every time step. While the algorithm attempts to maintain each node's desired degree for main and casual partnerships, if there are no available partners when a relationship dissolves, those nodes must wait until another relationship dissolves to form new partnerships. Table \ref{TableS1_rewiring} shows the average percentage of rewirings over 50 independent simulations, which do not happen in the same time step as the previous partnership dissolved. Even for the smallest network we examined, fewer than 3\% of the instances of rewiring could not occur immediately, demonstrating that this process nearly always occurs in the same time step, regardless of the overall size of the network. Therefore, we can be confident that the overall degree distribution for main and casual partnerships remains stationary.

\begin{table}[h!]
\caption{Percentage of Delayed Rewirings for Main and Casual Partnerships}\label{TableS1_rewiring}%
\begin{tabular}{@{}lll@{}}
\toprule
\textbf{Network Population Size} & \textbf{Main Partnerships}, Mean (SD) & \textbf{Casual Partnerships}, Mean (SD) \\
\midrule
N = 5,000   & 2.87 (0.83) & 2.54 $\times$ 10$^{-3}$ (0.014) \\[0.5ex]
N = 10,000  & 0.16 (0.14) & 1.68 $\times$ 10$^{-3}$ (5.73 $\times$ 10$^{-3}$) \\[0.5ex]
N = 20,000  & 2.70 $\times$ 10$^{-3}$ (0.01) & 2.16 $\times$ 10$^{-4}$ (1.51 $\times$ 10$^{-3}$) \\[0.5ex]
N = 40,000  & 0 (0) & 0 (0) \\[0.5ex]
N = 80,000  & 0 (0) & 0 (0) \\
\bottomrule
\end{tabular}
\end{table}

\begin{figure}[H]
\centering
\includegraphics[width = 1\textwidth]{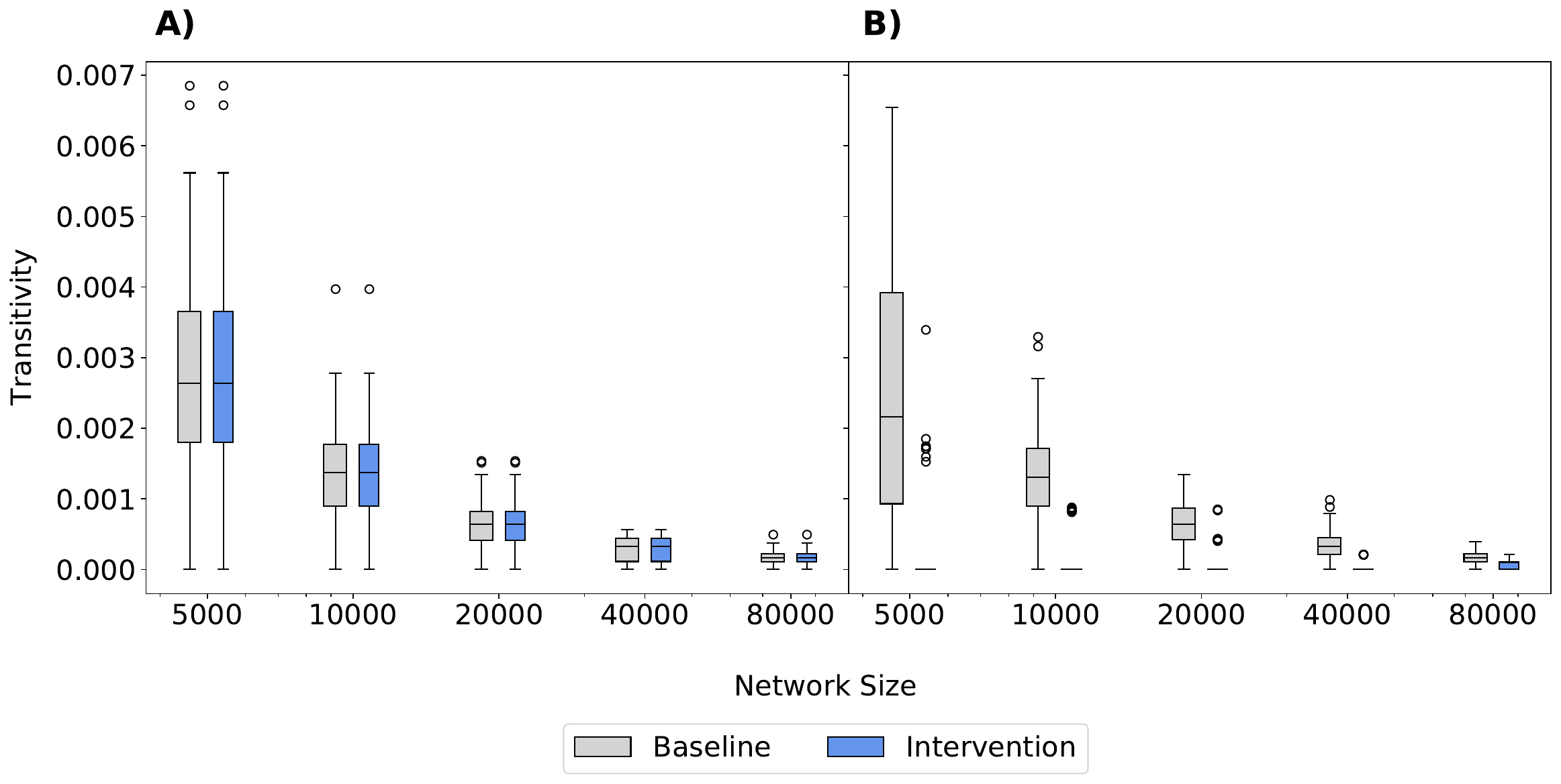}
\caption{\textbf{Comparison of Transitivity} Panels show the transitivity of the network at day 28 (Panel A) and day 84 (Panel B) of the simulation. Results presented are for 50 independent simulations of the baseline (no intervention) and intervention models.}\label{transitivity}
\end{figure}

\begin{figure}[H]
\centering
\includegraphics[width = 1\textwidth]{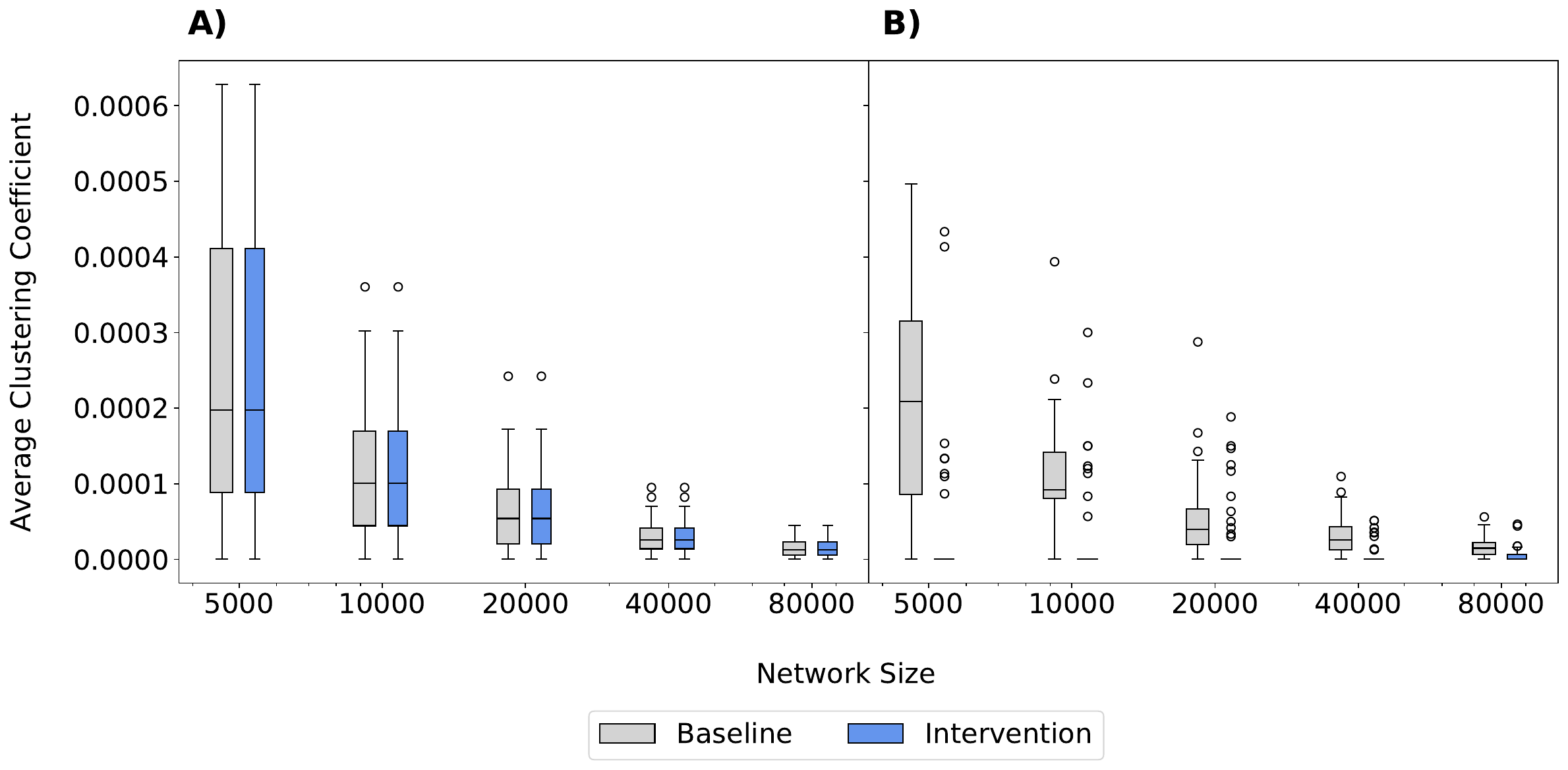}
\caption{\textbf{Comparison of Average Clustering Coefficient} Panels show the average clustering coefficient of the network at day 28 (Panel A) and day 84 (Panel B) of the simulation. Results presented are for 50 independent simulations of the baseline (no intervention) and intervention models.}\label{clustering}
\end{figure}

\begin{figure}[H]
\centering
\includegraphics[width = 1\textwidth]{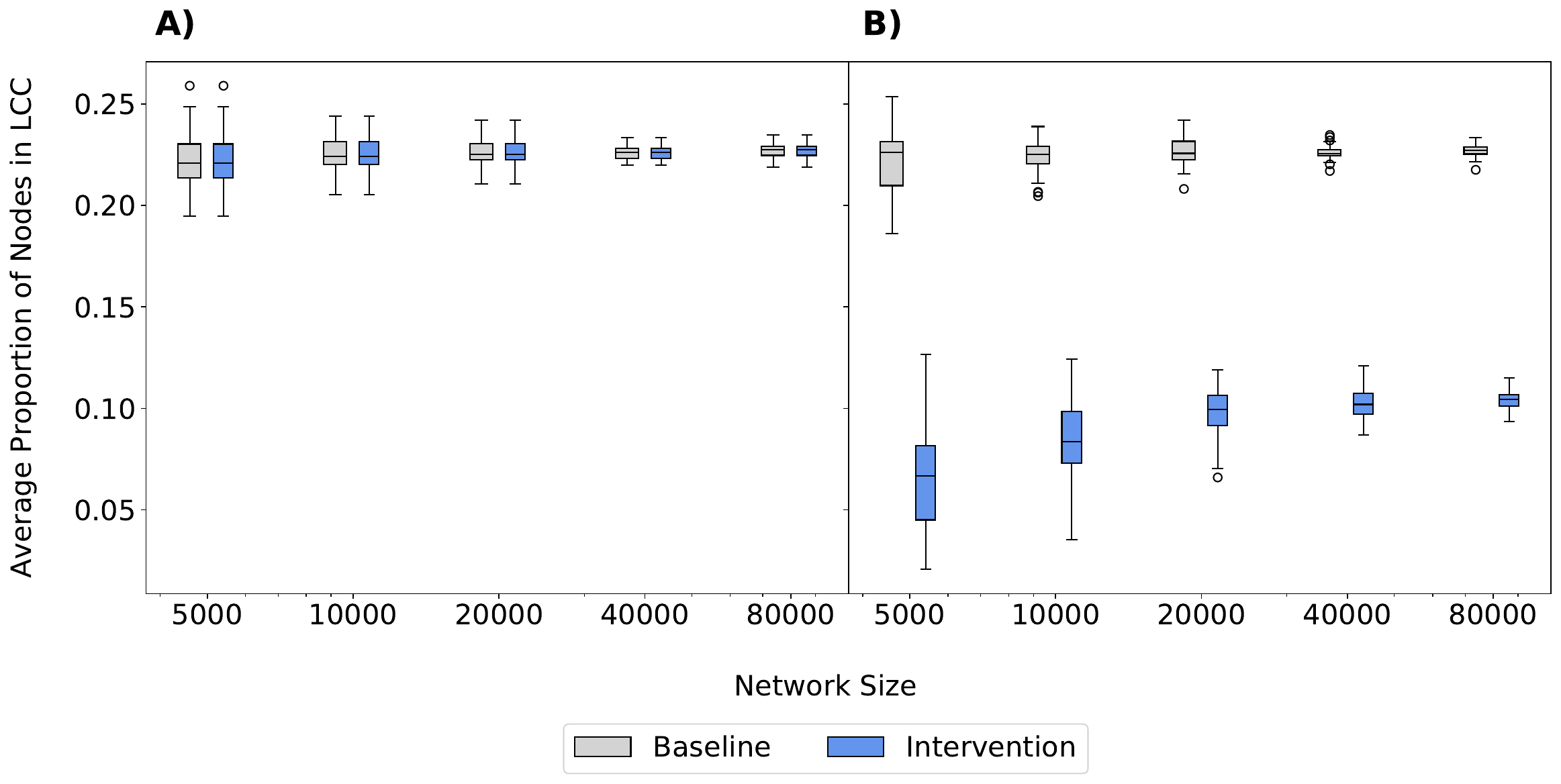}
\caption{\textbf{Comparison of the Proportion of Nodes Part of the Largest Connected Component} Panels show the proportion of nodes which are part of the largest connected component of the network at day 28 (Panel A) and day 84 (Panel B) of the simulation. Results presented are for 50 independent simulations of the baseline (no intervention) and intervention models.}\label{LCC}
\end{figure}

\begin{figure}[H]
\centering
\includegraphics[width = 1\textwidth]{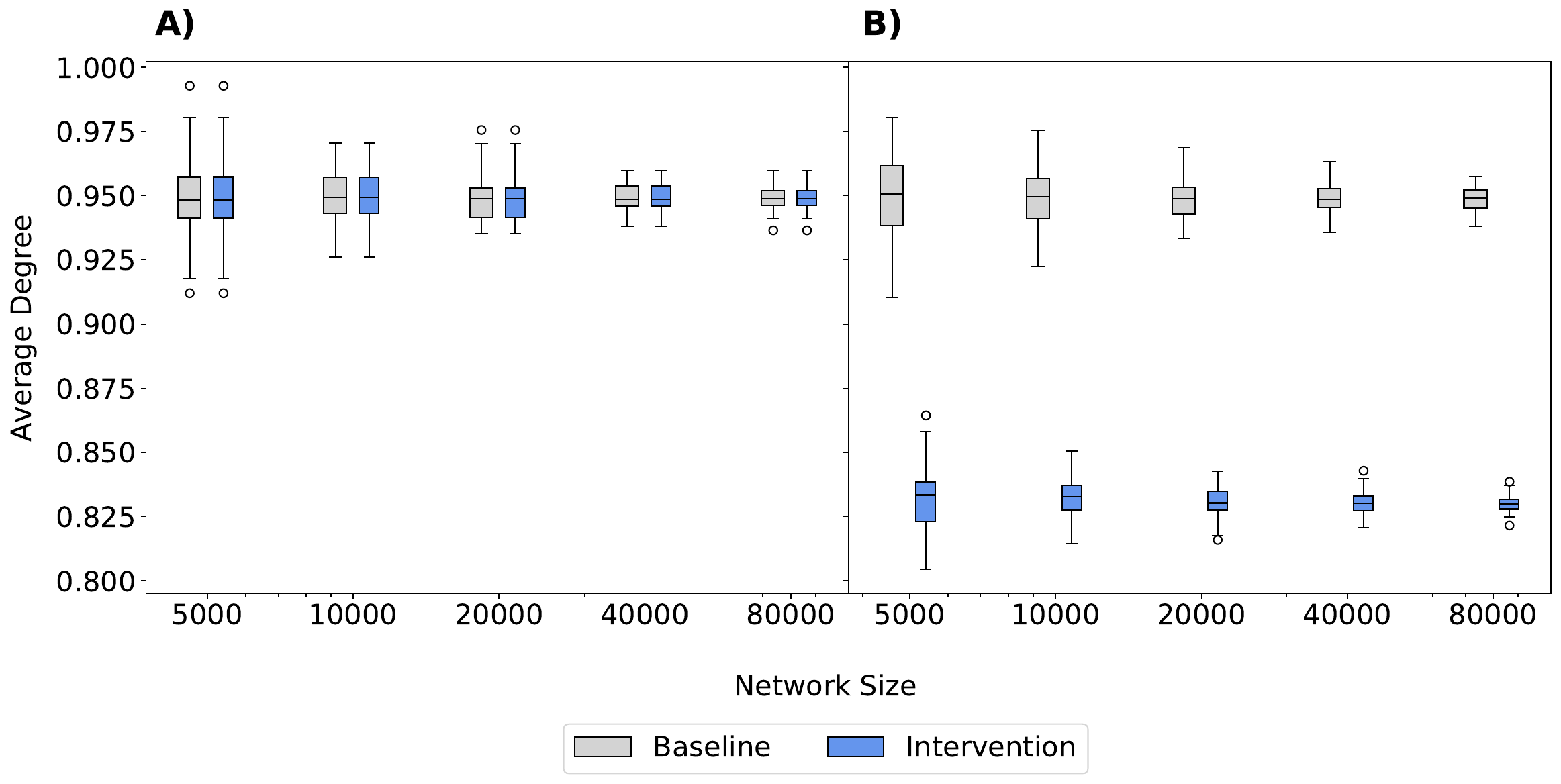}
\caption{\textbf{Comparison of Average Degree} Panels show the average degree of the network at day 28 (Panel A) and day 84 (Panel B) of the simulation. Results presented are for 50 independent simulations of the baseline (no intervention) and intervention models.}\label{avgdeg}
\end{figure}

\begin{figure}[H]
\centering
\includegraphics[width = 1\textwidth]{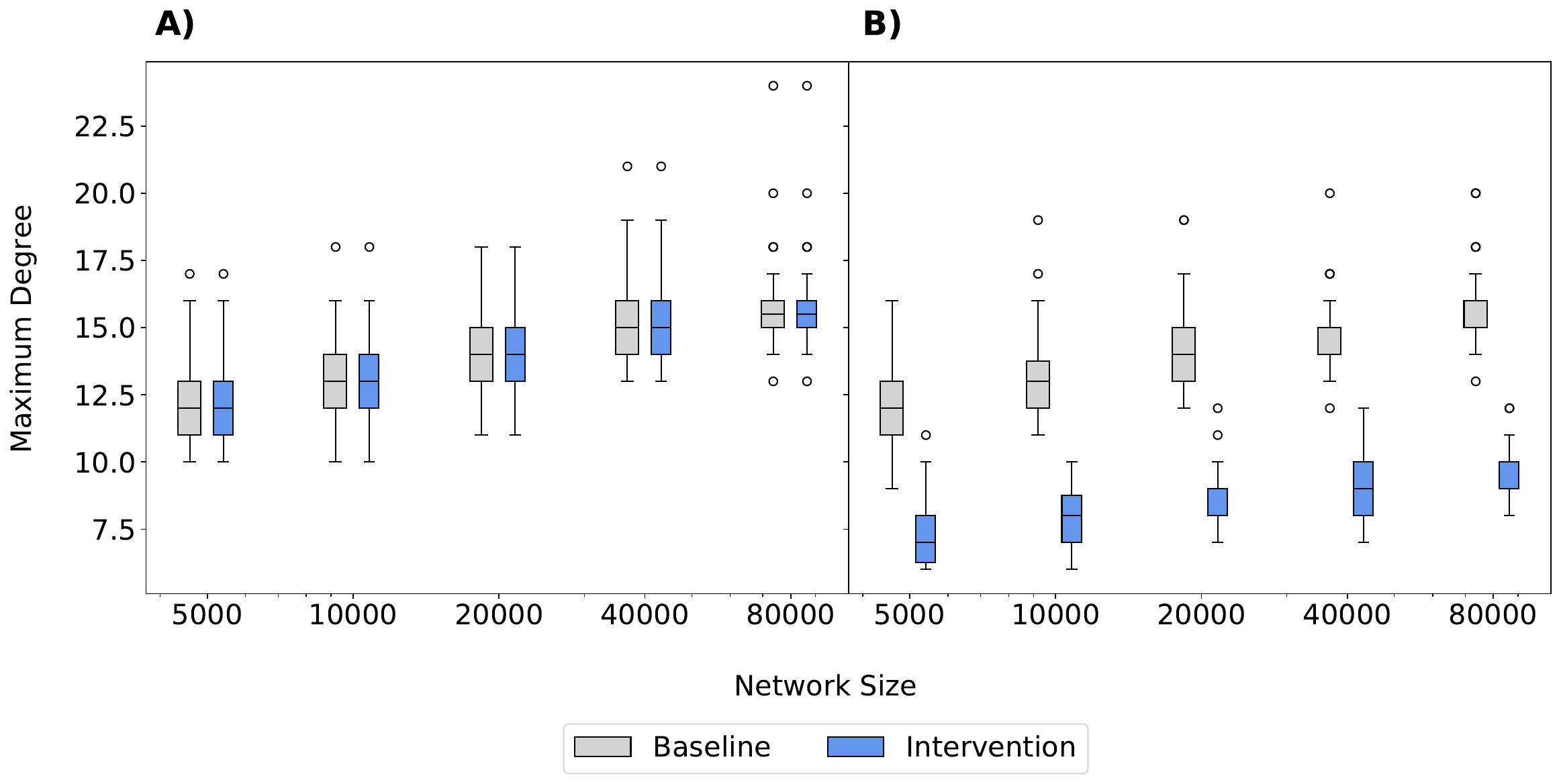}
\caption{\textbf{Comparison of Maximum Degree} Panels show the maximum degree of the network at day 28 (Panel A) and day 84 (Panel B) of the simulation. Results presented are for 50 independent simulations of the baseline (no intervention) and intervention models.}\label{maxdeg}
\end{figure}

To better understand the structure of the network, we examined different measures of network structure: transitivity, defined as the percentage of completed triangles (Figure \ref{transitivity}); average clustering coefficient (Figure \ref{clustering}); the proportion of nodes contained in the largest connected component (LCC) (Figure \ref{LCC}); average node degree (Figure \ref{avgdeg}); and maximum degree in the network (Figure \ref{maxdeg}). Given that the network is dynamic, we look at cumulative edges over a 7-day period in the network. We compare these network summaries for different population sizes, between baseline and intervention models, and at different times throughout the outbreak. The intervention compared is that of the main result, with vaccination beginning on day 30 and behavior change beginning on day 70. The timings selected were day 28, before any intervention, and day 84, two weeks after the second intervention in order to visualize the impact of intervention on network structure.

In general, we find that on day 28, prior to intervention, the baseline and intervention models are identical, which is as expected. After the intervention, summaries of network connectedness decrease. This is true in both the baseline and intervention models. In baseline models, this is due to the increased number of infected individuals as the infection spreads and, thus, the increased number of individuals beginning to comply with isolation recommendations. In the intervention scenarios, the decrease in network connectedness is further pronounced by implementing behavioral changes.

Average transitivity, or the proportion of completed triangles in the graph, as well as the average clustering coefficient, the fraction of completed possible triangles involving a particular node and then averaged over all nodes, can be thought of as measures of local connectivity in the graph. In this context, they look at how frequently two partners have a partnership with the same third partner. We expect this to decline with increased network size when partnerships are defined by random chance, as in the case of this model, because there are more nodes with which to partner.

The proportion of nodes contained in the LCC corresponds to a time-averaged dyanmic graph. In static graphs, the proportion of nodes in the LCC can be a rough indicator of the maximum proportion of nodes who could be infected by a disease-spreading process from a single source. Given that our network model is dynamic, this can be considered only as a snapshot in time and could be highly variable if, for example, a new edge forms between two connected components. However, the change in the proportion of nodes in the LCC post-intervention is still informative; it indicates an overall decline in the global connectedness of the graph, which indicates that it is less likely for an infection to become widespread over the network.

Average degree and maximum degree are summaries of node-level connectedness in the graph. In Figure \ref{avgdeg}, we see that, on average, the typical individual in our model has fewer than one partner over a single week. As expected, this highlights how sparse the network is. Maximum degree looks at the most connected node in the graph; in a disease-spreading process, this person can be considered a 'super spreader' and is often someone whose behavior greatly impacts the outbreak. While we expect the maximum degree in a graph to increase slightly with the overall size of the graph, this is somewhat limited. The number of one-time partners an individual seeks on a particular day is defined by a geometric distribution, parameterized by their daily probability of having a one-time partner, which is unaffected by network size. It is important to highlight the change in maximum degree post-intervention; in this context, it demonstrates that the maximum number of nodes that could be infected by a single node over a week has declined.

\subsection{Simulation Algorithms}

\begin{algorithm}[H]
\caption{Graph Initialization}
\label{alg:create__graph}
\begin{algorithmic}[1]
\Require Number of nodes $N$, Daily probability of one-time partnership formation $\pi_{o,k}$, Relationship type distribution $rt_k$

\Repeat 

    \State Assign relationship type for each node: 
    \[
    rt_k[x] \sim \text{Categorical}([0.471, 0.167, 0.074, 0.22, 0.047, 0.021])
    \]
    \State Compute degree sequence of main and casual partners: 
    \[
    n_m[x] = \begin{cases}
        0, & \text{if } rt_k[x] \in \{0,1,2\} \\
        rt_k[x] - 2, & \text{otherwise}
    \end{cases}
    \]

    \[
    n_c[x] = \begin{cases}
        0, & \text{if } rt_k[x] \in \{0,3\} \\
        1, & \text{if } rt_k[x] \in \{1,4\} \\
        2, & \text{otherwise}
    \end{cases}
    \]
    
    \State Assign sexual activity stratum: 
    \[
    strat_{o,k} \sim \text{Categorical}([0.19, 0.19, 0.19, 0.19, 0.19, 0.05])
    \]
    \State Compute one-time degree sequence given sexual activity stratum:
    \[
    n_o[x] \sim \text{Geometric}(1 - \pi_{o,k}[strat_{o,k}])
    \]
   
\Until{$\sum n_m$, $\sum n_c$, and $\sum n_o$ are all are even}
\State Initialize graph $G$ with $N$ nodes
\State \Return $G, n_m, n_c, n_o, rt_k, strat_{o,k}$

\end{algorithmic}
\end{algorithm}

\begin{algorithm}[H]
\caption{Initialize Relationships in the Network}
\label{alg:init_relationships}
\begin{algorithmic}[1]
\Require Graph $G$, degree sequence sequences for one-time ($n_o$) main ($n_m$) and casual partnerships ($n_c$), average duration of main partnerships $rd_{m,e}$, and average duration of casual partnerships $rd_{c,e}$
\State Create stub lists for main $D_m$, casual $D_c$, and one-time partnerships $D_o$
\For{$x$ in $G$}
    \State append $n_m[x] \cdot x$ to $D_m$
    \State append $n_c[x] \cdot x$ to $D_c$
    \State append $n_o[x] \cdot x$ to $D_o$
\EndFor
\State Randomly shuffle $D_o$
\For{each pair in $D_o$}
    \State Create one-time relationship $(u,v)$
\EndFor
\State Randomly shuffle $D_m$
\For{each pair in stub list of $D_m$}
    \State Create main relationship $(u,v)$ with duration $\sim \text{Geometric}(rd_{m,e})$
\EndFor
\State Randomly shuffle $D_c$
\For{each pair in stub list of $D_c$}
    \State Create casual relationship $(u,v)$ with duration $\sim \text{Geometric}(rd_{c,e})$
\EndFor
\State \Return $G$, main relationships, casual relationships, one-time relationships
\end{algorithmic}
\end{algorithm}

\begin{algorithm}[H]
\caption{Update One-Time Relationships}
\label{alg:update_onetime}
\begin{algorithmic}[1]
\Require Graph $G$, One-time encounter probability $\pi_{o,k}$, Activity stratum $\pi_{o,k}$
\State Remove all existing one-time relationships
\Repeat
    \For{each node $x \in G$}
        \State Sample number of one-time partners: 
        \[
        n_o[x] \sim \text{Geometric}(1 - \pi_{o,k}[x])
        \]
    \EndFor
\Until{Total stubs $\sum n_o$ is even}
\State Shuffle one-time stubs and randomly pair nodes to form relationships
\For{each pair $(u,v)$}
    \State Add one-time edge $(u,v)$
\EndFor
\State \Return $G, n_o$
\end{algorithmic}
\end{algorithm}

\begin{algorithm}[H]
\caption{Update Main and Casual Relationships}
\label{alg:update_relationships}
\begin{algorithmic}[1]
\Require Graph $G$

\State Remove expired relationships:
\For{each edge $(u,v)$ in $G$}
    \If{Relationship duration has ended}
        \State Remove edge $(u,v)$
        \If{main relationship}
            \State Add $u,v$ to list of nodes wanting a main partner, $W_m$, update $X_m[u] = v$, $X_m[v] = u$
        \ElsIf{casual relationship}
            \State Add $u,v$ to list of nodes wanting a casual partner, $W_c$, update $X_c[u] = v$, $X_c[v] = u$
        \EndIf
    \EndIf
\EndFor
\State Form new main partnerships:
\While{$|W_m| > 1$}
    \State Select node $x \in W_m$, find partner $y \notin X_m[x]$
    \If{$y$ exists}
        \State Create main relationship $(x,y)$ with duration $\sim \text{Geometric}(rd_{m,e})$
        \State Remove $x,y$ from $W_m$
    \EndIf
\EndWhile
\State Form new casual partnerships:
\While{$|W_c| > 1$}
    \State Select node $x \in W_c$, find partner $y \notin X_c[x]$
    \If{$y$ exists}
        \State Create casual relationship $(x,y)$ with duration $\sim \text{Geometric}(rd_{c,e})$
        \State Remove $x,y$ from $W_c$
    \EndIf
\EndWhile
\State \Return Updated $G$
\end{algorithmic}
\end{algorithm}

\begin{algorithm}[H]
\caption{Infection Spread in the Network}
\label{alg:spread}
\begin{algorithmic}[1]
\Require Graph $G$, Probability of infection $\beta$, Susceptible nodes $S$, Exposed nodes $E$, Infected nodes $I$, Contact probabilities $\pi_m, \pi_c, \pi_o$, Vaccine efficacy $VE$

\For{each infected node $i$ in $I$}
    \State Determine compliance level for isolation
    \State Identify susceptible neighbors $N_i = \{ j \mid j \in S \}$
    \For{each $j \in N_i$}
        \State Determine contact probability $\pi_m, \pi_c, \pi_o$ based on relationship type
        \If{Contact occurs and transmission succeeds $(U \sim \text{Uniform}(0,1) < \beta \cdot VE_j)$}
            \State Add $j$ to $E$ and remove $j$ from $S$
            \State Record infection source and time
        \EndIf
    \EndFor
\EndFor
\State \Return $S$, $E$, $I$, infection sources and time
\end{algorithmic}
\end{algorithm}

\begin{algorithm}[H]
\caption{Update Infection Status}
\label{alg:update_status}
\begin{algorithmic}[1]
\Require  Susceptible nodes $S$, Exposed nodes $E$, Infected nodes $I$, Recovered nodes $R$, Infection times $t_{i,k}$, Exposure times $t_{e,k}$, Treatment delays $T_d$, Current step $t$
\State Update remaining time in exposed and infected states:
\For{each $x \in E$} \quad $t_{e,k}[x] \gets t_{e,k}[x] - 1$
\EndFor
\For{each $x \in I$} \quad $t_{i,k}[x] \gets t_{i,k}[x] - 1$
\EndFor
\State Identify individuals transitioning states
\State\hspace{\algorithmicindent} $E \to I$: $E_{\text{new}} \gets \{ x \mid t_{e,k}[x] \leq 0 \}$
\State\hspace{\algorithmicindent} $I \to R$: $I_{\text{new}} \gets \{ x \mid t_{i,k}[x] \leq 0 \}$
\State \Return $S, E, I, R$
\end{algorithmic}
\end{algorithm}

\begin{algorithm}[H]
\caption{Vaccination Process}
\label{alg:vaccination}
\begin{algorithmic}[1]
\Require Graph $G$, Vaccine availability at time $t$ for each dose $V_{1,t}, V_{2,t}$, Vaccination probabilities for each dose $\pi_{v_1}, \pi_{v_2}$, Vaccine efficacy for each dose $VE_1, VE_2$
\State Identify eligible nodes for first dose: $W_1 = \{ x \notin I \bigcup R, V[x] = 0 \}$
\State Allocate vaccines: $V_1 \gets \text{Sample}(W_1, \min(W_1, V_{1,t}))$
\For{each $x \in V_1$}
    \State Set first dose received $(V[x] = 1)$
\EndFor
\State Identify eligible nodes for second dose: $W_2 = \{ x \mid V[x] = 1, \text{Time elapsed} \geq 28 \}$
\State Allocate second doses: $V_2 \gets \text{Sample}(W_2, \min(W_2, V_{2,t}))$
\For{each $x \in V_2$}
    \State Set second dose received $(V[x] = 2)$
\EndFor
\State Update vaccine efficacy: $VE[x] = VE_2$ if $V[x] = 2$, else $VE_1$
\State \Return Updated vaccine status
\end{algorithmic}
\end{algorithm}

\begin{algorithm}[H]
\caption{Run Simulation}
\label{alg:simulate}
\begin{algorithmic}[1]
\Require Number of nodes $N$, Initial infections $n_{init}$, Infection probability $\beta$, Steps $T$, Intervention start $T_{int}$, Behavior change $\pi_b$, Isolation level $iso$, Vaccination delay $T_v$, Daily vaccine availability over time $V_1, V_2$
\State Initialize network $G$ with $N$ nodes using \textsc{CreateGraph}
\State Initialize infected nodes $I$
\State Generate exposure times $t_{e,k} \sim \mathcal{N}(7,1)$ and infection times $t_{i,k} \sim \mathcal{N}(27,3)$
\State Compute initial relationship structure using \textsc{init\_relationships}
\For{each time step $t = 1$ to $T$}
    \If{No exposed or infected nodes remain}
        \State \textbf{break}
    \EndIf
    \If{$t = T_{int}$}
        \State Apply behavior change $\pi_b$
    \EndIf
    \If{$t \geq T_v$}
        \State Administer vaccinations using \textsc{vaccinate}
    \EndIf
    \State Update infection status using \textsc{update\_status}
    \State Spread infection using \textsc{spread}
    \State Update relationships using \textsc{update\_onetime} and \textsc{update\_relationships}
    \State Record epidemic dynamics: $S, E, I, R$
\EndFor
\State \Return $S, E, I, R$
\end{algorithmic}
\end{algorithm}

\end{appendices}


\section{References}



  \end{document}